\documentclass[a4size,journal,twoside]{IEEEtran}

\usepackage{amsmath,amsfonts}
\usepackage{algorithmic}
\usepackage{algorithm}
\usepackage{array}
\usepackage[caption=false,font=normalsize,labelfont=sf,textfont=sf]{subfig}
\usepackage{textcomp}
\usepackage{stfloats}
\usepackage{url}
\usepackage{verbatim}
\usepackage{graphicx}
\usepackage{cite}
\usepackage{xcolor}
\usepackage{hyperref}
\usepackage{tabularray}
\usepackage{svg}
\usepackage{amssymb}
\usepackage{pifont}
\usepackage{hyperref}
\newcommand{\cmark}{\ding{51}}%
\newcommand{\xmark}{\ding{55}}%

\usepackage{booktabs}
\usepackage{multirow}

\usepackage{cleveref}[2012/02/15]

\usepackage[T1]{fontenc}

\definecolor{light-gray}{gray}{0.75}

\newcommand{\figref}[1]{Figure~\ref{#1}}
\crefformat{footnote}{#2\footnotemark[#1]#3}

\begin{document}

\title{BASICS: Broad Quality Assessment of Static Point Clouds in a Compression Scenario}

%








\author{Ali Ak,~\IEEEmembership{Member,~IEEE,}
        Emin Zerman, 
        Maurice Quach, 
        Aladine Chetouani,~\IEEEmembership{Senior Member,~IEEE,}
        Aljosa Smolic, 
        Giuseppe Valenzise,~\IEEEmembership{Senior Member,~IEEE,}
        Patrick Le Callet,~\IEEEmembership{Fellow,~IEEE}
\thanks{A. Ak and P. Le Callet are with Nantes Universit\'e, \'Ecole Centrale Nantes, CNRS, LS2N, UMR 6004, F-44000 Nantes, France. 

P. Le Callet is with Institut Universitaire de France (IUF).

E. Zerman is with Department of Computer and Electrical Engineering, Mid Sweden University, Sundsvall, Sweden.

M. Quach was with Université Paris-Saclay, CNRS, CentraleSupélec, Laboratoire des Signaux et Systèmes (UMR 8506), Gif-sur-Yvette, France during the majority of the work, and he is now with Bosch Center for Artificial Intelligence, Renningen, Germany.

A. Chetouani is with Laboratoire PRISME, Universit\'e d'Orl\'eans, Orl\'eans, France.

A. Smolic is with Lucerne University of Applied Sciences and Arts (HSLU), Rotkreuz, Switzerland.

G. Valenzise is with Université Paris-Saclay, CNRS, CentraleSupélec, Laboratoire des Signaux et Systèmes (UMR 8506), Gif-sur-Yvette, France.

}
}

\markboth{Author Accepted Manuscript, January~2024}%
{\textit{A}\MakeLowercase{\textit{k et al.}}: Author Manuscript}

\IEEEoverridecommandlockouts
\IEEEpubid{\makebox[\columnwidth]{Published at IEEE TMM - DOI: \href{https://dx.doi.org/10.1109/TMM.2024.3355642}{10.1109/TMM.2024.3355642} - 
\copyright IEEE \hfill} \hspace{\columnsep}\makebox[\columnwidth]{ }}

\maketitle

\begin{abstract}
Point clouds have become increasingly prevalent in representing 3D scenes within virtual environments, alongside 3D meshes. Their ease of capture has facilitated a wide array of applications on mobile devices, from smartphones to autonomous vehicles. Notably, point cloud compression has reached an advanced stage and has been standardized. However, the availability of quality assessment datasets, which are essential for developing improved objective quality metrics, remains limited. In this paper, we introduce BASICS, a large-scale quality assessment dataset tailored for static point clouds. The BASICS dataset comprises 75 unique point clouds, each compressed with four different algorithms including a learning-based method, resulting in the evaluation of nearly 1500 point clouds by 3500 unique participants. Furthermore, we conduct a comprehensive analysis of the gathered data, benchmark existing point cloud quality assessment metrics and identify their limitations. By publicly releasing the BASICS dataset, we lay the foundation for addressing these limitations and fostering the development of more precise quality metrics.
\end{abstract}


\section{Introduction}
\label{sec:intro}

\IEEEPARstart{D}{igital} imaging technologies have revolutionized the capability to capture real-world environments and recreate them in different temporal or spatial contexts. This capacity has extended to the realm of 3D scenes through the integration of computer graphics and photogrammetry techniques. Presently, we can capture intricate 3D objects and scenes using an array of tools, ranging from solely RGB cameras to RGB cameras supplemented with additional sensors~\cite{collet2015high, zhou2021tightly, schreer2019capture, roynard2018paris, rosnell2012point}. Within the domain of 3D modeling, two primary representations have gained prominence: colored point clouds and textured 3D meshes~\cite{zerman2020vsensevvdb2}. 
In this paper, our focus is directed toward point clouds, a representation widely used in numerous applications, in particular augmented and virtual reality.
The acquisition of point clouds is typically accomplished through diverse means, including stereo-camera arrays~\cite{schreer2019capture}, LiDAR sensors~\cite{roynard2018paris} and conventional cameras~\cite{rosnell2012point}. Point cloud acquisition produces huge amounts of data, calling for compression techniques for efficient transmission and storage.

However, evaluating the performances of compression algorithms is time consuming and expensive. While the field of research in this area has witnessed notable expansion in recent years\cite{zerman2020vsensevvdb2, perry2020pcqadset, dasilvacruz2019pcqatowardsdefinition, yang2021projbasedPCQAdset, su2019PCQAdset, Hua2020VQACPCAN, wu2021siatpcqd, subramanyam2020vvqdset, alexiou2020pointxr}, there are still many open questions and problems.

Existing datasets in the field have notable shortcomings in terms of diversity, scale, consistency, and data accessibility. These limitations constitute substantial challenges for research into learning-based approaches. The absence of diversity is frequently attributed to several factors, including the recurrent use of the same point clouds across various datasets, limited geometric complexity and semantic categories, as well as the utilization of the same compression algorithms as the sole source of distortion types. This lack of diversity is further compounded by the relatively modest scale of the existing datasets, rendering them ill-suited for the development of learning-based quality metrics. Consistency issues also can be observed in existing datasets, stemming from factors such as the utilization of unnormalized point clouds and inconsistent rendering specifications. A detailed overview of the existing datasets is presented in Table \ref{tab:DatasetOverview}, which summarizes the aforementioned deficiencies. Another substantial shortcoming is the limited data availability and incompleteness, e.g., due to copyright constraints or repository management practices, as well as the absence of individual opinion scores, standard deviations and confidence intervals, which are typically needed to develop more sophisticated metrics. The accessibility of data in existing datasets is summarized in Table \ref{tab:public_availability}.

In conclusion, the need for a new dataset is evident due to aforementioned deficiencies in existing datasets. Our proposal seeks to address these shortcomings comprehensively, providing a dataset that encompasses all essential characteristics required for the development of more accurate quality metrics. Furthermore, it provides further insights into the performance of point cloud quality metrics when applied to point clouds encoded with learning-based compression algorithms.

The contributions of this work are threefold:
\begin{itemize}
    \item We present, and make publicly available, a broad point cloud quality assessment dataset featuring 75 unique point clouds that hold semantic relevance within the context of telepresence scenarios (as detailed in Section~\ref{sec:database}).
    \item We compare the performances of various state-of-the-art methods for point cloud compression (as outlined in Section~\ref{sec:subjective_experiment_results}).
    \item We provide an exhaustive  benchmark of the state-of-the-art point cloud quality metrics, including both point-based and image-based assessment (as outlined in Section~\ref{sec:objective}).
\end{itemize}

The BASICS dataset has been made publicly available\footnote{\label{fn:zenodo}BASICS Dataset Link: https://zenodo.org/doi/10.5281/zenodo.8324545
} under the Creative Commons Attribution-NonCommercial-ShareAlike (CC BY-NC-SA 4.0) license with the aim of fostering continued research within the field. 
The repository contains individual and mean opinion scores, as well as pristine and compressed point clouds. Additionally, the dataset's GitHub page\footnote{GitHub: https://github.com/kyillene/BASICS-Public} provides the scripts required for the evaluation of point cloud quality metrics.
The BASICS dataset has been successfully employed in the grand challenge on point cloud quality assessment\footnote{https://sites.google.com/view/icip2023-pcvqa-grand-challenge/} organized by some of the authors at the 2023 IEEE International Conference on Image Processing (ICIP2023).

\begin{table*}[]
\caption{Statistical summary of existing PC quality assessment datasets}
\label{tab:DatasetOverview}
\resizebox{\textwidth}{!}{%
\begin{tabular}{@{}lcccccllll@{}}
\toprule
\multicolumn{1}{c}{Dataset}                                      & nb SRC & nb PPC & \begin{tabular}[c]{@{}c@{}}nb obs\\ per PPC\end{tabular} & \begin{tabular}[c]{@{}c@{}}total \\ nb obs\end{tabular} & Display & \multicolumn{1}{c}{Visualization} & \multicolumn{1}{c}{\begin{tabular}[c]{@{}c@{}}Subj. test\\ method\end{tabular}} & \multicolumn{1}{c}{\begin{tabular}[c]{@{}c@{}}Temporal\\ dimension\end{tabular}} & \multicolumn{1}{c}{Distortions}                                                                                          \\ \midrule
BASICS (proposed)                                                & 75     & 1494   & 60                                                       & 3600                                                    & 2D      & Passive                           & DSIS (side-by-side)                                                             & Static                                                                           & Compression: GPCC, VPCC, GEOCNN                                                                                          \\
vsenseVVDB2\cite{zerman2020vsensevvdb2}                          & 8      & 136    & 23                                                       & 23                                                      & 2D      & Passive                           & ACR                                                                             & Dynamic                                                                          & Compression: GPCC, VPCC                                                                                                  \\
Perry et al.~\cite{perry2020pcqadset}                            & 6      & 90     & -                                                        & 73                                                      & 2D      & Passive                           & DSIS (side-by-side)                                                             & Static                                                                           & Compression: GPCC, VPCC                                                                                                  \\
da Silva Cruz et al.~\cite{dasilvacruz2019pcqatowardsdefinition} & 8      & 48     & -                                                        & 50                                                      & 2D      & Passive                           & DSIS                                                                            & Static                                                                           & \begin{tabular}[c]{@{}l@{}}Octree pruning\\ Projection-based compression from 3DTK\end{tabular}                          \\
Yang et al.~\cite{yang2021projbasedPCQAdset}                     & 10     & 420    & 16                                                       & 64                                                      & 2D      & Passive                           & ACR                                                                             & Static                                                                           & \begin{tabular}[c]{@{}l@{}}Octree pruning, random point down-sample\\ Color noise, geometric gaussian noise\end{tabular} \\
Su et al.~\cite{su2019PCQAdset}                                  & 20     & 740    & -                                                        & 60                                                      & 2D      & Passive                           & DSIS (side-by-side)                                                             & Static                                                                           & \begin{tabular}[c]{@{}l@{}}Octree pruning, geometric gaussian noise\\ Compression: SPCC, VPCC, LPCC\end{tabular}         \\
NBU-PCD1.0~\cite{Hua2020VQACPCAN}                                & 10     & 160    & -                                                        & -                                                       & 2D      & -                                 & -                                                                               & Static                                                                           & Octree pruning                                                                                                           \\
SIAT-PCQD~\cite{wu2021siatpcqd}                                   & 20     & 340    & 38                                                       & 76                                                      & HMD     & Interactive                       & DSIS                                                                            & Static                                                                           & Compression: VPCC                                                                                                        \\
Subramanyam et al.~\cite{subramanyam2020vvqdset}                 & 8      & 64     & -                                                        & 52                                                      & HMD     & Interactive                       & ACR-HR                                                                          & Dynamic                                                                          & Compression: VPCC                                                                                                        \\
PointXR~\cite{alexiou2020pointxr}                                 & 5      & 40     & 20                                                       & 40                                                      & HMD     & Interactive                       & DSIS                                                                            & Static                                                                           & Compression: GPCC                                                                                                        \\ \bottomrule
\end{tabular}%
}
\end{table*}

\begin{table}[]
\caption{Public availability of the existing datasets}
\label{tab:public_availability}
\resizebox{\columnwidth}{!}{%
\begin{tabular}{lcc}
\hline
Dataset                                                          & Point Clouds         & Subjective Annotations \\ \hline
BASICS (proposed)                                                & \cmark               & Individual Scores      \\
vsenseVVDB2~\cite{zerman2020vsensevvdb2}                         & \cmark               & Individual Scores      \\
Perry et al.~\cite{perry2020pcqadset}                            & Broken URL           & \xmark                 \\
Su et al.~\cite{su2019PCQAdset}                                  & \cmark               & (D)MOS only            \\
da Silva Cruz et al.~\cite{dasilvacruz2019pcqatowardsdefinition} & \xmark               & \xmark                 \\
Yang et al.~\cite{yang2021projbasedPCQAdset}                     & \cmark               & (D)MOS only            \\
NBU-PCD1.0~\cite{Hua2020VQACPCAN}                                & \xmark               & \xmark                      \\
SIAT-PCQD~\cite{wu2021siatpcqd}                                  & Broken URL           & \xmark                 \\
Subramanyam et al.~\cite{subramanyam2020vvqdset}                 & \xmark               & Individual Scores      \\
PointXR~\cite{alexiou2020pointxr}                                & \cmark               & Individual Scores      \\ \hline
\end{tabular}%
}
\end{table}

\section{The BASICS dataset}
\label{sec:database}

The BASICS dataset has been designed and collected with two main objectives in mind: on one hand, providing diverse and extensive data to train learning-based point cloud quality assessment metrics; and on the other hand, creating challenging test conditions to benchmark existing quality metrics.
In this section, we describe the various stages of the dataset generation process.

\subsection{Material selection}
\label{subsec:database_materials}

\begin{figure*}[!t]
    \centering
    \includegraphics[width=\textwidth]{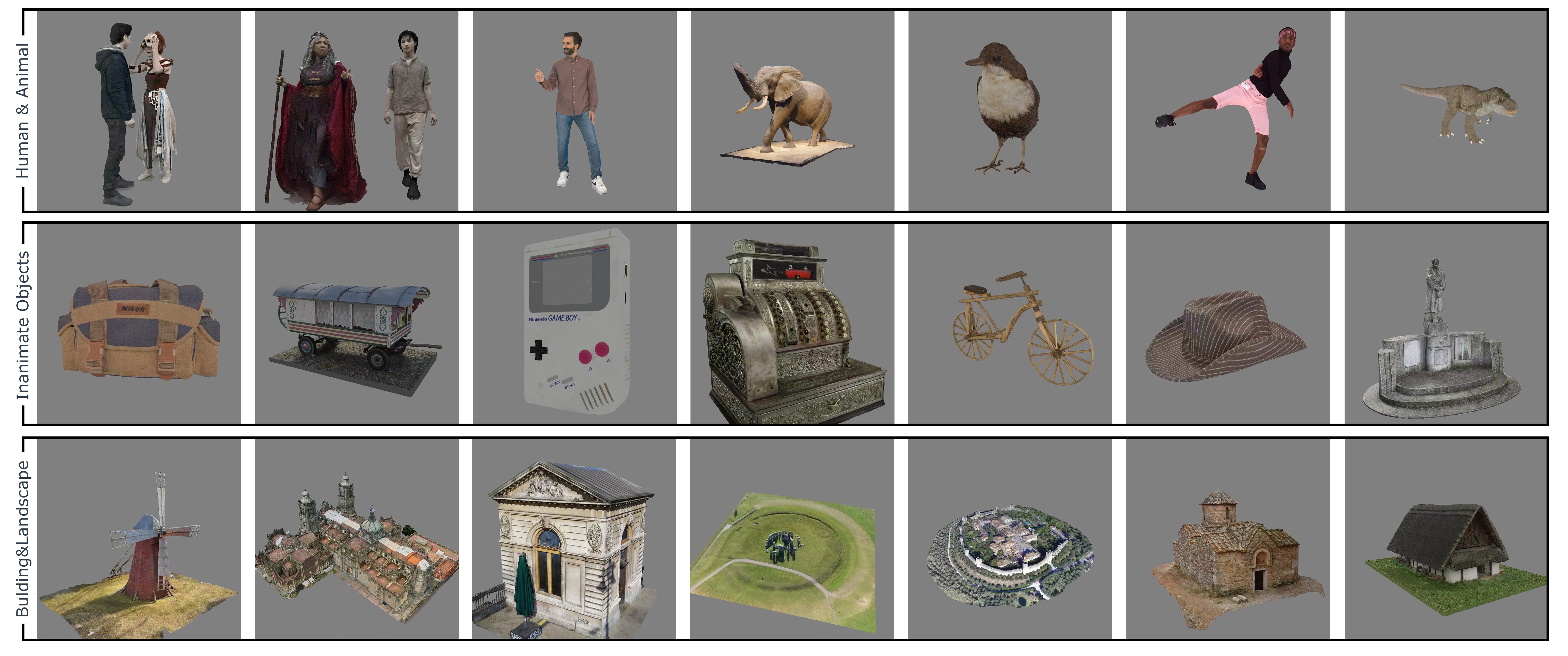}
    \caption{Sample renderings of 7 point clouds from each of the 3 semantic categories in the dataset (i.e., 7 of the total of 25 for each category).}
    \label{fig:semantic_cat_sample_frames}
\end{figure*}

In order to construct a point cloud quality assessment dataset that encompasses semantic diversity tailored for telepresence applications, we gathered a total of 75 point clouds, distributed across three semantic categories. Despite the broad scope of these categories, they remain highly relevant within the context of telepresence applications. The categories we have defined are ``Humans \& Animals (HA)'', ``Inanimate Objects (IA)'', ``Buildings \& Landscapes (BL)'', collectively covering a comprehensive set of subjects such as animals, humans, everyday items, vehicles, architectural structures and natural landscapes. \figref{fig:semantic_cat_sample_frames} offers a visual glimpse into each of these categories.

For the data collection and material selection part, our primary objective was to collect a comprehensive repository of publicly available point clouds that could be freely redistributed. However, numerous data sources impose strict restrictions on redistribution. Consequently, we acquired the 3D models from two sources: collaborator studios (i.e., V-SENSE studio\footnote{https://v-sense.scss.tcd.ie/} and XD Productions\footnote{https://www.xdprod.com/}) as well as an online repository for 3D model sharing called SketchFab\footnote{https://www.sketchfab.com/}. Even in this case, availability of point clouds was somewhat limited. To address this, we gathered 3D meshes and generated point clouds via sampling the mesh surface, as detailed in Section~\ref{subsec:database_preprocess}.

In total, 104 models were handpicked by three authors of this paper considering the semantic categories described above. Following the removal of materials that exhibited significant similarity, models with highly reflective surfaces and imperfect texturing, as well as those with loose semantic associations to their respective categories, the total count of models was reduced to a final selection of 75. For each model, information regarding its source (a SketchFab URL if applicable), file format, statistics, and licensing details can be accessed in the dataset repository\cref{fn:zenodo}.

\subsection{Pre-processing}
\label{subsec:database_preprocess}

In order to mitigate potential distortions stemming from other sources, the collected 3D models require a pre-processing and a conversion into point clouds before any further processing. This was necessary due to the heterogeneous formats in which these models were originally available. The scripts used for pre-processing are provided as a toolbox and available on Github\footnote{https://github.com/mauriceqch/2023\_BASICS\_PC\_toolbox}. 

Models that were already in point cloud PLY format required minimal attention, except for voxelization. Conversely, 3D meshes underwent through several steps. These steps are further discussed below. 

\subsubsection{Making 3D meshes uniform}

The collected 3D mesh models were in various formats. All 3D meshes were converted into OBJ format, to streamline the pre-processing chain, using Blender\footnote{https://www.blender.org/} and Meshlab\footnote{https://www.meshlab.net/}. 

\subsubsection{Cleaning 3D meshes}
Some of the 3D meshes had either parts that had transparent or reflective properties (e.g., glasses in some models), which could not be replicated well during point cloud rendering. Some other meshes had incomplete parts (e.g., trees, some of the building fa\c{c}ades) which would decrease the users' quality of experience and introduce other sources of distraction and distortion. To avoid such effects, these parts were removed or cleaned in Blender.

The mesh files were then unified into a single OBJ file, so that the sampling process in the pipeline could be done with ease. Next, the material properties (which are described in the .mtl files) are checked to eliminate any other reflective properties of the materials, which could not be reproduced correctly in the point cloud format. After all these operations, the 3D meshes were ready for the point cloud sampling step.

\subsubsection{Sampling point clouds}
Using CloudCompare\footnote{https://www.cloudcompare.org/main.html}, point clouds were sampled from the 3D meshes' surfaces. During this operation 10 million points were sampled on the surfaces of the said meshes. The sampling operation extracted the location, color, and normal attributes for each point in the point clouds. At the end of this stage, all 3D models were in (originally or after conversion) point cloud format.

\subsubsection{Point cloud voxelization}

We used 10-bit quantization for point cloud voxelization, as higher bit-depths showed no notable benefits.
That is, the spatial coordinates are normalized such that they are integers between 0 and 1023. Only unique points remained after voxelization step.
This has two main advantages: first, the coordinates are in a range that is predictable for point cloud processing but also with respect to rendering and second, we use voxelized coordinates in combination with cube based rendering to improve stability, predictability and quality of renderings.

\subsection{Compression}
\label{subsec:database_compression}

\begin{figure*}[!t]
    \centering
    \includegraphics[width=\textwidth]{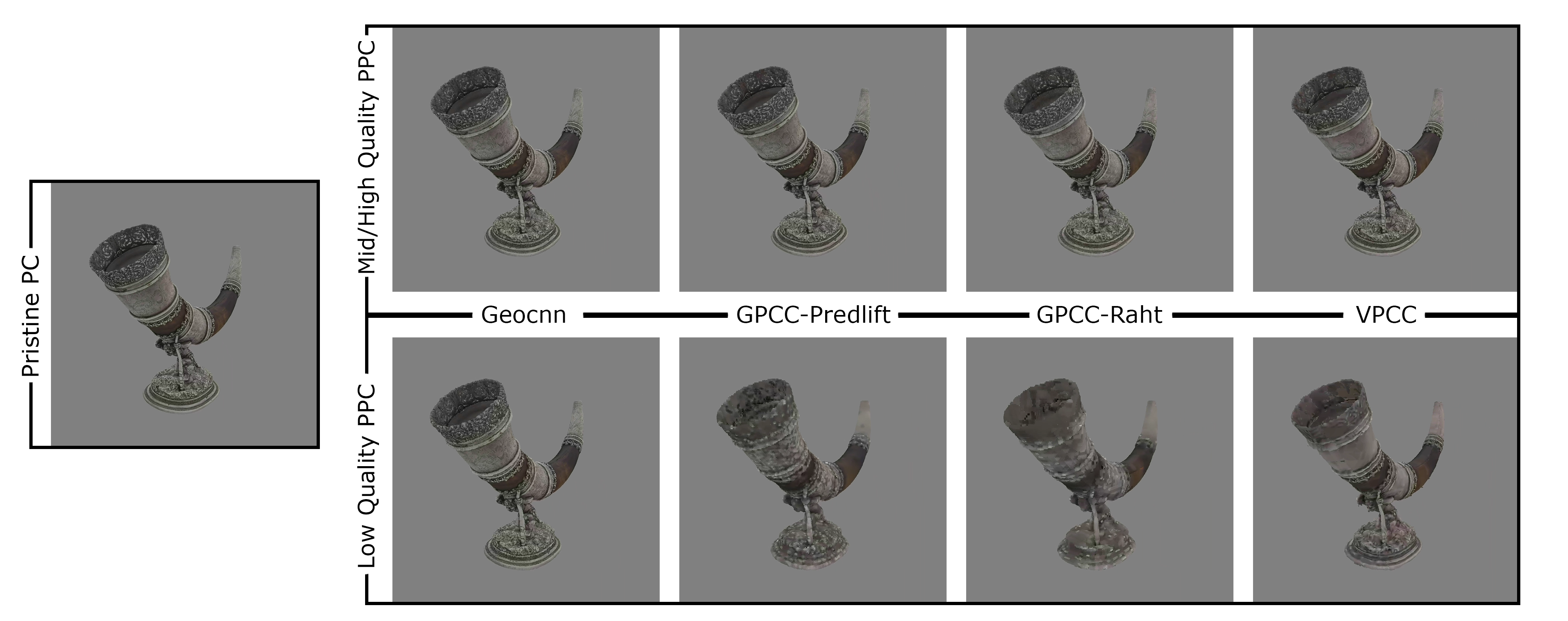}
    \caption{Sample frames from the video renderings of a selected processed point cloud (PPC), showing results for each compression algorithm. 
    }
    \label{fig:compression_samples}
\end{figure*}

Compression is a crucial stage for various point cloud storage and transmission applications, including streaming and telepresence. It is also the most realistic distortion source for point clouds. In this context, one needs to compress point clouds to transfer them to the remote receiver under current bandwidth limitations. 

We selected four compression algorithms: 
\begin{itemize}
    \item GPCC-Octree-Predlift (noted as GPCC-Predlift),
    \item GPCC-Octree-RAHT (noted as GPCC-Raht), 
    \item VPCC, and
    \item GeoCNN.
\end{itemize} 
VPCC and GPCC were selected as they are part of MPEG standardization efforts, and they are among the most commonly used compression methods. GeoCNN was selected to introduce artifacts of a learning-based compression to BASICS database. Sample results are shown in \figref{fig:compression_samples}.

GPCC~\cite{graziosi2020overview} compresses the point cloud using the octree structure which can find the occupancy of the points in 3D space without projection to 2D space. GPCC uses either octree or trisoup (Triangle soup) approaches -- based on a pruned octree. As octree and trisoup only focus on geometry compression, attributes (such as color) are compressed using region-adaptive hierarchical transform (RAHT) and predicting/lifting transform (indicated as pred/lift or predlift).

For GPCC, it was noticed during the pilot tests that the trisoup generated uneven structure and holes on the reconstructed point clouds. Therefore, in this database, only octree is used for GPCC geometry coding, together with the two attribute coding methods: Octree-Predlift and Octree-Raht. For compression, the six quality parameter (QP) values from common test conditions (CTC)~\cite{schwarz2018mpeg_ctc} were used. The worst quality level (i.e., r1) was discarded in the pilot tests because its quality was too bad and this would affect the subjective quality experiment adversely by changing the rating scale and the participants' votes. With this change, the number of quality levels for GPCC became five: $r_{GPCC} \in \{2, 3, 4, 5, 6\}$.

VPCC~\cite{graziosi2020overview} is the video-based point cloud compression approach, in which the point cloud is projected to the sides of a cube and the projection is coded using traditional video compression methods, such as HEVC/H.265 or VVC/H.266. The projection is done for both color information and also the depth information (i.e., the inherent 3D structure of points in 3D). Utilizing the inherent temporal coding capabilities of traditional video codecs, VPCC can effectively compress dynamic point cloud sequences. 

For VPCC, we used the ``all-intra'' mode, and the compression levels were taken from the ``longdress'' sequence QP as determined in the CTC. It should be noted that the compression levels were not optimized for each content, as this was out of scope of this study.

Nevertheless, it was noticed that the given bitrates in the CTC seemed to yield a much higher quality than GPCC, therefore, another quality level below the ones given in CTC was added. This quality level is called \textit{rate\textunderscore{}zero} with $gQP=36$ and $tQP=47$. With this change, the number of quality levels for VPCC became six: $r_{VPCC} \in \{0, 1, 2, 3, 4, 5\}$.

GeoCNN~\cite{quach2020geocnn} compresses voxelized point clouds by first performing block partitioning.
Then, each block is passed to a variational autoencoder where the encoder transforms the input binary occupancy voxel grid to a latent space.
The latent space is then quantized and entropy coded using a learned entropy model.
After entropy decoding the bitstream, the latent space is transformed back to a voxel grid containing predicted occupancy probabilities.
The probabilities are then thresholded to binary values which yields the decoded block.
With the result of each block, the entire decompressed point cloud is obtained. Four different quality levels were used for GeoCNN. $r_{GeoCNN} \in \{1, 2, 3, 4\}$.

\section{Subjective quality assessment}
\label{sec:subj_experiment}

We conducted a large-scale subjective experiment using the Prolific~\cite{prolific} crowdsourcing platform with 3654 participants. 1494 processed point clouds (PPC) from 75 original point clouds (SRC) were generated for the experiment with each PPC were evaluated by 60 unique participants on average.  Consequently, we successfully accumulated approximately 90,000 subjective opinion scores. This section describes the details regarding the crowdsourcing study.

\subsection{Generating Visual Stimuli}
\label{subsec:subjective_render}

In a voxelized point cloud, a voxel is occupied if at least one point of the point cloud is within the voxel. Each voxel is rendered as a cube spanning its volume. This is different from the more common point-based method (``point'' as OpenGL point primitive) where points are rendered as screen-aligned squares of a given window space size. The main drawback of the point-based rendering is its susceptibility to scaling issues: as the viewer zooms in or out, the points can either become smaller or larger, impacting the perceived density of the point cloud. Furthermore, point-based rendering often results in flicker artifacts, particularly when spatial overlaps occur during perspective changes. Cube-based rendering addresses these issues, ensuring consistent rendering quality across various perspectives and effectively eliminating flicker artifacts.

In practice, we specify cube sizes.
For octree-based methods, the size of the cube is defined based on the number of removed octree levels $n_r$.
With $n_l$ bit quantization, the maximum is $n_l$ levels.
Thus, we specify the size of the cube as $2^{n_l - n_r}$: that is, a size of 1 when lossless, a size of 2 when removing one octree level, a size of 4 when removing two octree levels, etc\ldots For other methods, the cube size is determined empirically to ensure water-tight rendering.

\begin{figure}[!t]
    \centering
    \includegraphics[width=\columnwidth]{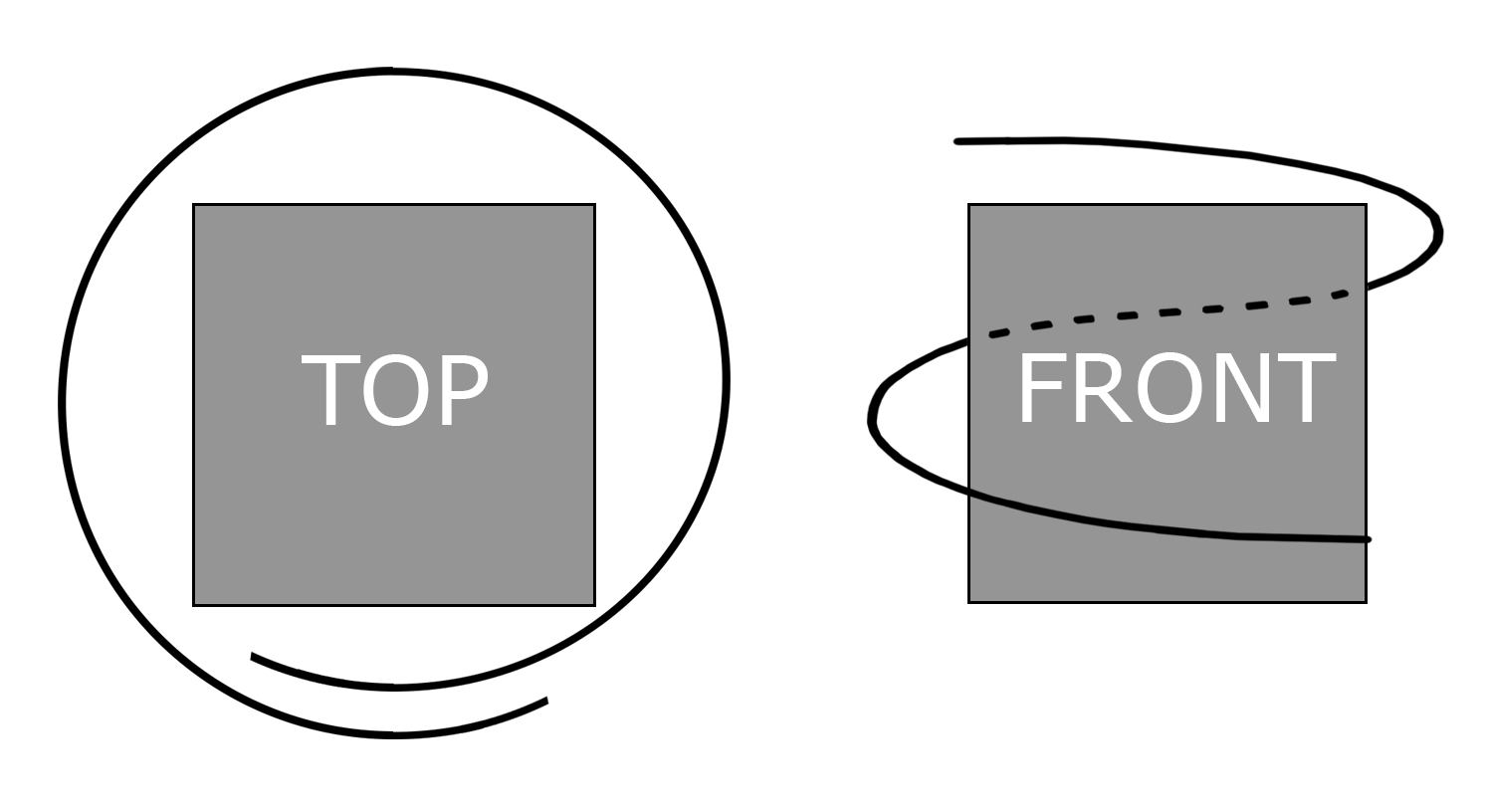}
    \caption{Visualization of the rendering trajectory from top and front views.}
    \label{fig:rendering_trajectory}
\end{figure}

We employed a helix-like rendering trajectory, as illustrated in \figref{fig:rendering_trajectory}. The frontal orientation of each point cloud was manually designated, and the rendering trajectory consistently started from this predefined front side. To ensure that the front side of the point cloud remained visible either at the beginning or at the end of the rendered video, we introduced a slight overlap between the trajectory's starting and ending points. Certain point clouds are unnatural to observe from lower angles (e.g., landscape, buildings). Therefore, each point cloud was individually categorized as either ``low'', ``mid'' or ``high'' to determine the initial elevation of the rendering trajectory. While moving on the rendering trajectory, the camera was always oriented toward the center of the point cloud.

\subsection{Methodology}
\label{subsec:subjective_methodology}

\begin{figure}[!t]
    \centering
    \includegraphics[width=\columnwidth]{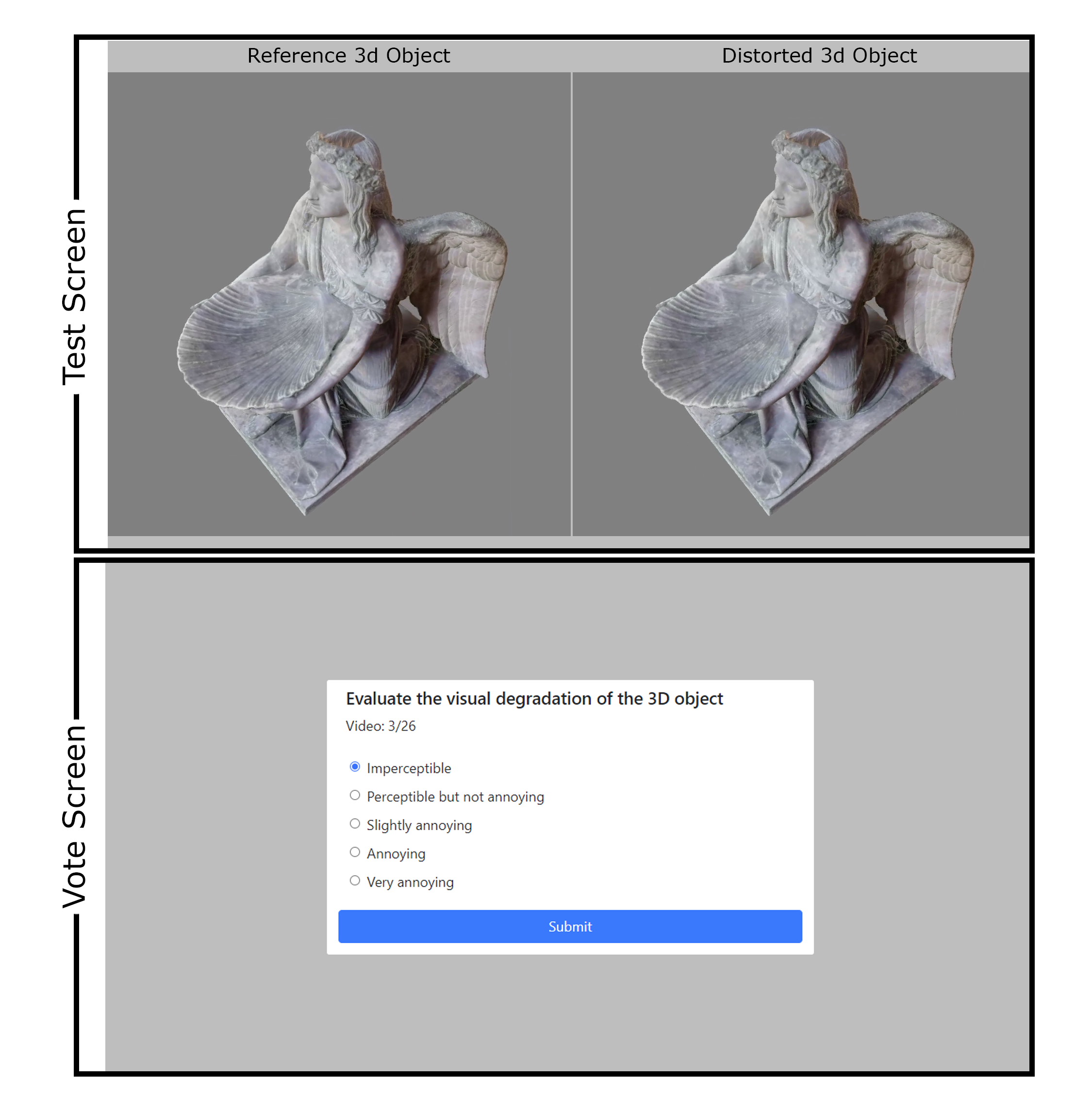}
    \caption{Sample screenshots from the experiment. Rendered point cloud videos were shown side-by-side (above), and each stimulus was followed by a voting screen (below).}
    \label{fig:sample_test_screen}
\end{figure}

Subjective quality assessment of point cloud content can be broadly categorized as interactive and passive~\cite{valensize2023book_subj_obj_vv}. In the interactive paradigm, observers have the liberty to examine the point cloud from any desired point of view, often within the context of augmented reality or virtual reality applications. In contrast, the passive approach involves rendering point clouds in the form of a conventional video with a predetermined camera trajectory. Although both paradigms have their own advantages and drawbacks, there is no statistically significant difference between the subjective opinions collected with each approach~\cite{vioal2022interactiveVSpassive}. In order to minimize the variance between observer opinions and facilitate a more practical data collection through crowdsourcing, we opted for the passive approach~\cite{nehme2021crowdsourcingreliabilityMeshes}.

Several methodologies can be found in the literature and recommendations for subjective quality assessment of traditional image and video sequences~\cite{itu2019BT500}. Commonly employed methodologies include, but are not limited to, Absolute Category Rating (ACR), Double Stimulus Impairment Scale (DSIS), Two-Alternative Forced Choice (2AFC). Several studies compared the accuracy and reliability of each methodology for diverse types of multimedia content. 
In the realm of traditional images and videos, it is shown that the pair comparison methodology tends to be more accurate due to straightforward experiment procedure and there is no statistically significant difference between ACR and DSIS methodology~\cite{mantiuk2012SubjMethodComp2D}. However, it's worth noting that the pair comparison methodology may become impractical when dealing with a substantial number of test conditions due to the exponential increase in required comparisons~\cite{zerman2018pcCrosscontent}. 
On the other hand, the recent study by Nehme \textit{et al}.~\cite{nehme2019SubjMethodComparison} suggests that the DSIS method is more accurate than ACR for 3D graphical content. The rationale behind this assertion is that in ACR experiments, participants unfamiliar with the pristine models may struggle to discern various types of distortions. DSIS methodology enhances accuracy by presenting both the reference and the distorted model before the rating phase, allowing for a more informed evaluation. In line with these findings, we adopted the DSIS methodology with a side-by-side presentation format, as recommended by Nehme \textit{et al.} \cite{nehme2019SubjMethodComparison}. A screenshot depicting the stimuli presented to the participants is featured in \figref{fig:sample_test_screen}.

\subsection{Test Procedure}
\label{subsec:subjective_procedure}

Previous studies have indicated that crowdsourcing experiments can yield results of comparable accuracy to traditional laboratory experiments across various Quality of Experience (QoE) tasks and with diverse experiment designs~\cite{goswami2021crowdsourcingreliabilityTMO, nehme2021crowdsourcingreliabilityMeshes}. In light of these findings, we chose to leverage the Prolific crowdsourcing platform to recruit participants and conduct our subjective experiment.
Prolific ensures transparency and ethical participation by clearly communicating to participants that they are taking part in a research study. The experiment requirements are thoughtfully balanced to benefit both researchers and participants alike~\cite{palan2018prolificmainpaper}. This approach not only provides access to a broad pool of participants but also expedites the data collection process.

\begin{figure}[!t]
    \centering
    \includegraphics[width=\columnwidth]{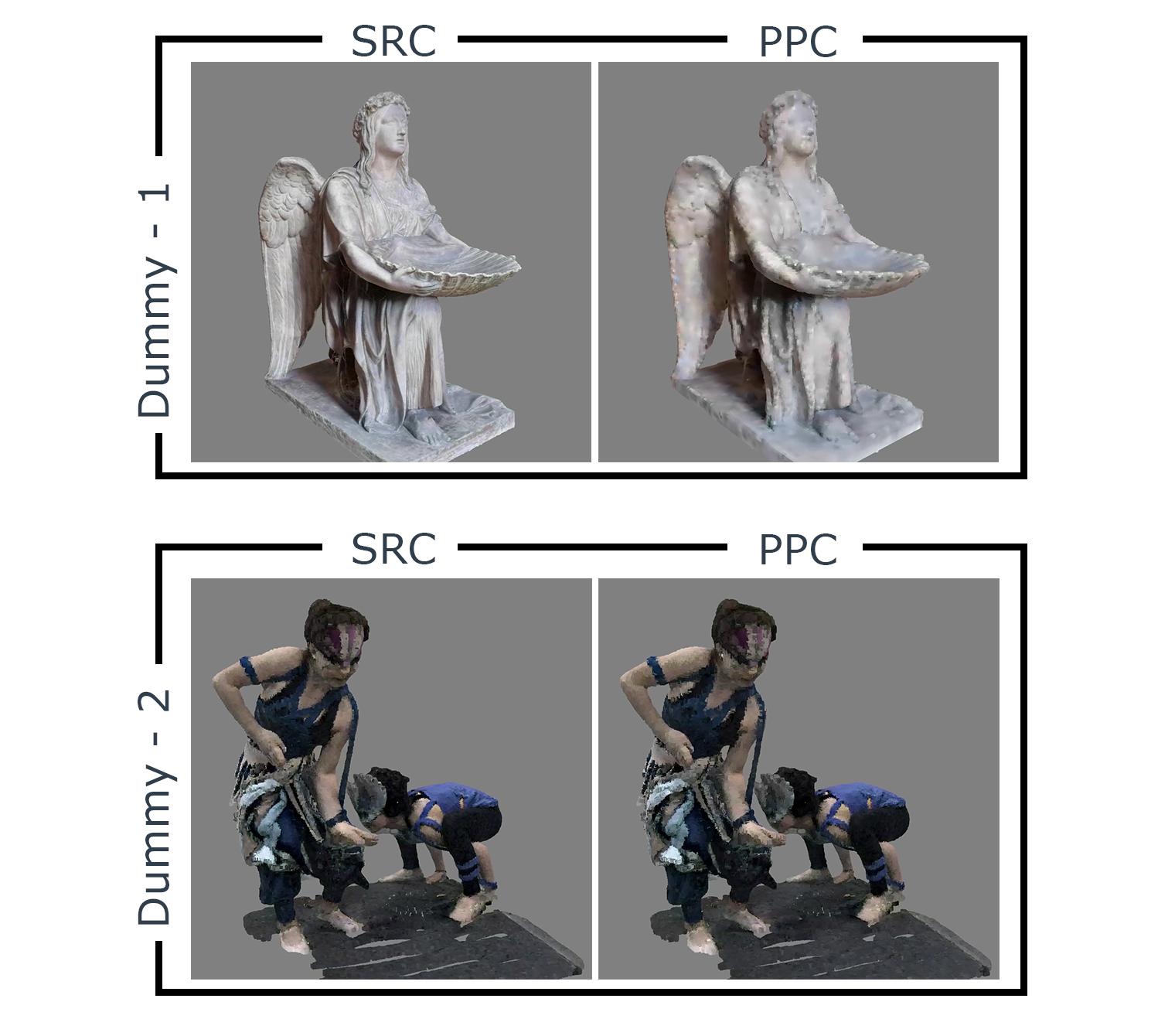}
    \caption{Selected frames from the SRC and PPC video renderings of the 2 dummies used in all playlists.}
    \label{fig:dummy_screenshots}
\end{figure}

\textbf{Test sessions \& Duration:} In crowdsourcing settings where participants lack supervision during the experiment, it is essential to limit both the number of stimuli and the duration of the test compared to laboratory experiments~\cite{goswami2021crowdsourcingreliabilityTMO}. To accommodate this requirement, we divided the experiment into 60 sessions, each containing 25 stimuli and 2 ``dummies''. Sample frames from the two dummy stimuli are presented in \figref{fig:dummy_screenshots}. For training purposes, one dummy from the highly compressed stimuli and one dummy from the least compressed stimuli were uniformly presented to every participant. These dummy stimuli remained consistent for all participants, and participants were not informed that these stimuli were included for training purposes. In total, each participant rated 27 stimuli of 10-seconds rendered videos. With unlimited voting time after each stimulus presentation, the average duration of the test sessions amounted to approximately 5 minutes and 30 seconds.

\textbf{Participants \& Requirements:} We recruited 60 participants (50\% female - 50\% male) on average per session, 3654 participants in total. The age of the participants range from 18 to 70. Each participant was compensated for their time and effort in line with the Prolific requirements. Moreover, to uphold the integrity of the experiment and guarantee that all stimuli were presented as intended, participants were constrained to use specific browsers operating in full-screen mode at a resolution of 1080p. Additionally, participants were required to meet specific qualifications, namely completing at least 200 submissions on the Prolific platform with a 100\% approval rate. These stringent prerequisites helped ensure the reliability and commitment of the participants.

\section{Subjective Experiment Results}
\label{sec:subjective_experiment_results}

This section presents our analysis on the collected subjective quality scores. In Section \ref{subsec:observer_screening}, we discuss the observer reliability and provide an overview of the results obtained from the observer screening tools we have applied. Additionally, in Section \ref{subsec:compression_algorithm_performance}, we present our findings on the performance of compression algorithms.

\subsection{Observer Screening}
\label{subsec:observer_screening}

In addition to the recruitment requirements of the participants (see Section \ref{subsec:subjective_procedure}), the ``dummy stimuli'' described above were also used as trap questions to detect participants engaging in malicious behavior. Moreover, post-experiment observer screening tools were employed to enhance the reliability of the subjective opinion scores. 

As described in Section \ref{sec:subj_experiment}, the dataset underwent evaluation by a total of 3,654 participants, divided into 60 smaller playlists, each receiving an average of 60 participants. Within each playlist, we included 2 dummy stimuli (depicted in \figref{fig:dummy_screenshots}) designed to calibrate participants' expectations regarding the extent of distortions present in the experiment. Participants were not informed that these initial 2 stimuli were for training purposes and evaluated them like regular stimuli. Source and processed point cloud renderings were displayed side-by-side like the rest of the stimuli in the experiment (see \figref{fig:sample_test_screen} for what is displayed to observers). We selected the first dummy with the highest level compression in the dataset (resulting in a low-quality PPC), and the second dummy with the lowest level of compression (resulting in a high-quality PPC). This deliberate selection allowed us to expose each participant to both ends of the quality scale, aligning their expectations regarding the extent of degradation. Notably, if a subject rated the first dummy as "imperceptible" due to its clear distortions, we excluded them from the experiment, removing a total of 22 subjects.

As post-experiment observer screening, we applied the two common ITU standards as well as a recently proposed model called ZREC\footnote{https://github.com/kyillene/ZREC}~\cite{itu2019BT500, itu2021BT913, zhu2023zrec}. Our objective in employing these methodologies was to assess the validity of the collected subjective opinion scores and establish reliable MOS. While the three methods differ fundamentally, their primary aim is to reduce confidence intervals (CI) by identifying and, if necessary, excluding outliers from the collected data. Although there is a high correlation between the MOS obtained through each method, the average 95\% CIs differ significantly.

Initially, we calculated the average 95\% CI by taking the mean of the 95\% CI of all stimuli without applying any post-screening methods, referred to as raw-MOS. Subsequently, we implemented the outlier rejection as recommended in ITU.BT500~\cite{itu2019BT500}. Notably, BT500 identified only 43 subjects as outliers out of the total 3,633 participants. As a result, the impact on both the MOS and the 95\% CI was minimal, with the 95\% CI (0.1982) remaining almost as high as the raw-MOS 95\% CI.
Conversely, we observed a more substantial shift in the acquired MOS when using P913-12.6 and ZREC, which yielded much lower confidence interval values. Specifically, we calculated an average 95\% CI of 0.1697 and 0.1463 for P913-12.6 and ZREC, respectively.
Based on these findings, we utilized the MOS acquired via ZREC due to the lower CI. To promote further investigation, we provide both the raw opinion scores and estimated MOS using ZREC in the public repository of the dataset.

\subsection{Performance of compression algorithms}
\label{subsec:compression_algorithm_performance}

\begin{figure}[!t]
    \centering
    \includegraphics[width=\columnwidth]{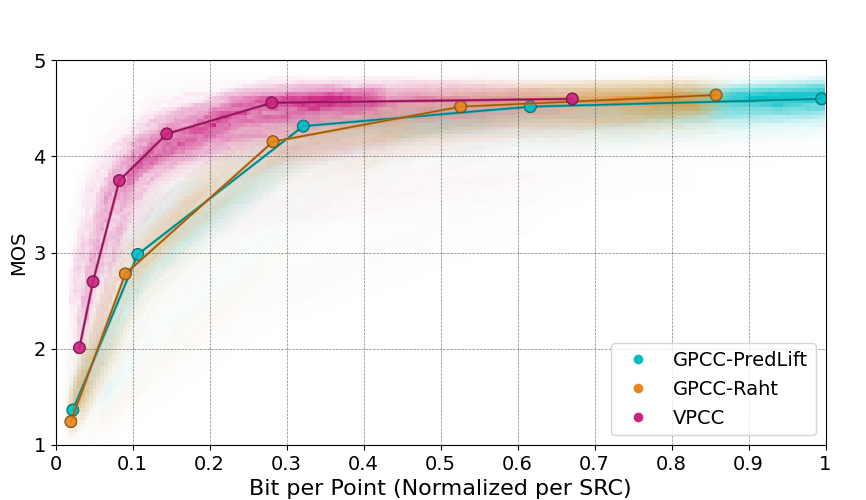}
    \caption{Heatmaps represent the MOS-BPP curve densities of each compression algorithm. Centers of the 2 dimensional Gaussian distribution fit over each compression level for each algorithm is plotted as a line.}
    \label{fig:rd_curve_density}
\end{figure}

Although BASICS dataset contains 4 compression algorithms, in this analysis, we excluded GeoCNN. Since GeoCNN only compresses geometry and is complemented with uncompressed color information for visualization, it is not fair to directly compare it with GPCC-Predlift, GPCC-Raht and VPCC, which code both geometry and color. 

When comparing the performance of GPCC-Predlift, GPCC-Raht and VPCC codecs, we employed two complementary approaches that analyze and visualize the MOS vs Bit-Per-Point (BPP) behavior of each compression algorithm. Due to large differences over BPP values per SRC, we first normalized all BPP values per SRC. We achieved this by dividing each PPC with the highest BPP value of the PPC across all compression algorithms that belongs to the corresponding SRC. Consequently, we could operate within a BPP range of 0 to 1, allowing a relative comparison of the compression algorithms. 

The first approach draws inspiration from DenseLines~\cite{moritz2018denselines}, a method originally devised to visualize vast quantities of time series data. To construct a Mean Opinion Score-Bit-Per-Point (MOS-BPP) curve density map for each compression algorithm, we partitioned the grid into square cells. Subsequently, for a given compression algorithm and SRC, we assigned a value of 1 to each cell intersecting with the polygon defined by the MOS and 95\% CI values of each PPC. For each SRC, we normalized the marked cell values by dividing them by the total number of intersecting cells, effectively illustrating the effect of lower 95\% CI and BPP values. Then, for each compression algorithms, we aggregated the normalized cell values from all SRCs and performed a linear mapping to scale the results into [0,1] range. Outputted values then plotted as overlaying heatmaps (with increasing transparency from 0 to 1), as showcased in \figref{fig:rd_curve_density}.

In the second approach, we employed a 2D Gaussian distribution to model each compression level within each compression algorithm. Subsequently, we extracted the centers of these 2D Gaussian distributions and utilized them to generate lines that captured the average MOS-BPP curves for each compression algorithm. These results were then plotted as lines superimposed upon the curve density heatmaps, providing a concise representation of each compression algorithm's performance.

As depicted in \figref{fig:rd_curve_density}, there is no significant difference between the GPCC-Predlift and GPCC-Raht. At lower BPP values, VPCC exhibits superior performance compared to both GPCC variants. However, as BPP values increase, the distinction between these compression algorithms becomes less pronounced. It's worth noting that while there may be occasional exceptions to these observations, they are minor in nature and do not substantially deviate from the trends observed in the analysis. We encourage interested readers to check the dataset public repository, where the MOS-BPP plots for each SRC is provided.

\section{Benchmark of Objective Quality Metrics}
\label{sec:objective}

Point cloud objective quality metrics can be categorized into three classes, based on the input to the metrics: image-based, color-based and geometry-based. Image-based metrics take the rendered point cloud image or image sequences as input and assess the quality of the point clouds. Geometry-based metrics primarily rely on the geometric information stored at each point in the point cloud, without considering color attributes. Color-based metrics, on the other hand, use the color information of each point to assess point cloud quality. Some metrics, such as PCQM~\cite{meynet2020pcqm}, have the capability to utilize both geometry and color information for quality assessment.

Furthermore, each metric can be categorized into three based on the presence of reference point cloud information as full-reference (FR), reduced-reference (RR) and no-reference (NR). FR metrics access all information from the reference point cloud in addition to the distorted point cloud. RR metrics can access only partial information (features) from the reference point clouds. NR metrics assess the quality of the point cloud without any access to the reference point cloud. 

In this section, we benchmark 14 image-based, 9 color-based and 17 geometry-based metrics from the literature. Some of these metrics were omitted from the results due to minor differences to their variants. An introduction to the selected metrics is provided in Section \ref{subsec:selected_metrics}. We employ various figures of merit and evaluation scenarios to assess the performance of these metrics, which are outlined in Section \ref{subsec:evaluation_criteria}. The results of the metric evaluations are presented in Section \ref{subsec:broad_quality_eval}, Section \ref{subsec:high_quality_eval}, and Section \ref{subsec:intra_src_eval}.

\subsection{Selected Metrics}
\label{subsec:selected_metrics}

For all image-based metrics, average pooling over 30 fps video renderings has been used to predict the final quality as recommended in \cite{ak2021temporalsampling}. Image-based metrics have been computed on the rendered frames used during the subjective test, and include simple measures such as MSE, PSNR, SSIM~\cite{wang2004ssim}, MS-SSIM~\cite{wang2003msssim}, and 11 other more sophisticated metrics. Feature Similarity Index (FSIM~\cite{zhang2011fsim}) and its color-dependent variant, FSIMc~\cite{zhang2011fsim}, fall under the category of full-reference metrics. They rely on phase congruency and gradient magnitude to locally quantify image quality, utilizing phase congruency as a weighting function to yield a single quality score. 
Gradient Magnitude Similarity Deviation (GMSD~\cite{xue2014gmsd}) is another full-reference metric that employs pixel-wise gradient magnitude similarity to predict image quality. 
D-JNDQ~\cite{ak2022djndq} is a learning-based full-reference metric trained on the first Just Noticeable Difference (JND) points of JPEG compression artifacts. It combines a white-box optical and retinal pathway model with a Siamese neural network to predict image quality.
MW-PSNR~\cite{stankovic2015mwpsnrFR, stankovic2016mwpsnrRR} relies on morphological wavelet decomposition and the Mean Squared Error (MSE) of the wavelet sub-bands. Our evaluation includes both full-reference (MW-PSNR-FR) and reduced-reference (MW-PSNR-RR) versions of this metric.
ADM2~\cite{li2011adm2} assesses image quality by separating detail losses and additive impairments. It encompasses features also used in the Video Multi-method Assessment Fusion (VMAF) metric~\cite{netflix2018VMAF}.
VIF~\cite{sheikh2006vif} quantifies the information present in the reference image and how much of this reference information can be extracted from the distorted image. It is another feature used in the VMAF metric~\cite{netflix2018VMAF}.
VMAF~\cite{netflix2018VMAF}, proposed by Netflix, fuses several image-based features, including ADP2 and VIF, along with a simple temporal feature to evaluate video quality.
FVVDP~\cite{mantiuk2021fvvdp} models the response of the human visual system to differences across the temporal domain and the visual field.

In addition to image-based metrics, several geometry-based metrics were also evaluated over the dataset. In the last decade, three fundamental approaches were proposed to evaluate PC quality focusing on point-to-point~\cite{mekuria2016point2point} (p2point), point-to-plane~\cite{tian2017point2plane} (p2plane) and plane-to-plane~\cite{alexiou2018plane2plane} (pl2plane) differences in 3D space. The p2point and p2plane metrics are computed using either mean square error (MSE) or peak peak signal-to-noise ratio (PSNR). In this context, the term ``plane'' refers to the surface of a point defined by its normal vector. The pl2plane metrics are computed using either MSE or root mean square (RMS). 
While categorized as a geometry-based metric, PCQM~\cite{meynet2020pcqm} uses a linear combination of several geometry-based (curvature comparison, curvature contrast, and curvature structure) and color-based features (lightness comparison, lightness contrast, lightness structure, chroma comparison, and hue comparison) to assess the visual quality of a point cloud. PointSSIM~\cite{evangelos2020pointssim} offers three geometry-based variants with slight differences in both implementation and performance. 

Moreover, color differences between the points in reference and distorted PC can be quantified with PSNR to estimate the visual quality of the distorted PC. We applied this metric on Y, U, and V channels separately and referred as Color-Y-PSNR, Color-U-PSNR, and Color-V-PSNR respectively. 3 color-based variants of the PointSSIM~\cite{evangelos2020pointssim} metric were also included in the evaluation.

\begin{figure}
    \centering
    \includegraphics[width=\columnwidth]{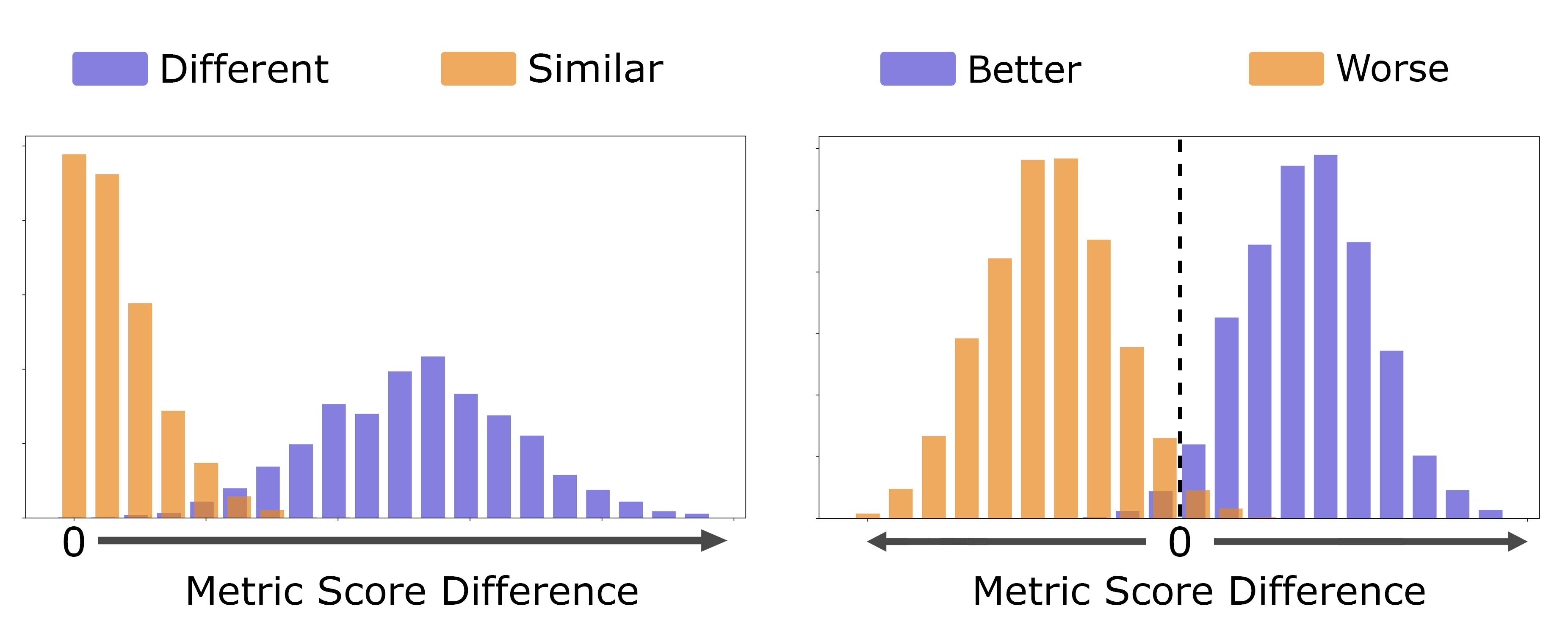}
    \caption{Ideal distributions of metric score differences for ``Different vs Similar'' and ``Better vs Worse'' analysis. A greater metric score difference is expected for different pairs in ``Different vs Similar'' analysis. For ``Better vs Worse'' analysis, metric score differences are expected to be positive and negative respectively for better and worse pairs.}
    \label{fig:ideal_distributions}
\end{figure}

\subsection{Evaluation Criteria}
\label{subsec:evaluation_criteria}

We evaluate the performance of the selected metrics in three different scenarios. In Section \ref{subsec:broad_quality_eval}, we aim to assess the perceptual quality of the PPC covering the \textit{entire quality range}. This is the most generic and traditional evaluation scenario for quality metrics. 
In Section \ref{subsec:high_quality_eval}, we evaluate the performance of the metrics in the \textit{high-quality range}. This scenario is particularly relevant for applications such as content production, high-quality streaming, and digital twins, where maintaining the highest visual fidelity is crucial.
In Section \ref{subsec:intra_src_eval}, we evaluate the metric performances for their \textit{sensitivity to quality differences} within different versions of the same point cloud content. This evaluation scenario is especially suitable to optimization scenarios such as point cloud compression and enhancement, and as loss functions in end-to-end PC learning pipelines. 
For broad and high-quality range evaluation scenarios, we employ traditional correlation measures, while Krasula's method~\cite{krasula2016metriceval} is used for the Intra-SRC evaluation scenario.

\textbf{Correlation measures:} Pearson's linear correlation coefficient (PLCC) measures the prediction accuracy of the objective metrics and Spearman's rank-order correlation coefficient (SROCC) measures the strength of prediction monotonicity~\cite{itu2019recP1401}. Following the recommendations~\cite{itu2019BT500, itu2019recP1401}, a 5-parameter logistic function was fitted prior to evaluation. For both PLCC and SROCC, the values are in the range [0, 1] and higher values indicate a better correlation.

\textbf{Krasula's method~\cite{krasula2016metriceval}:} It involves two stages of analysis: ``Different vs Similar'' and ``Better vs Worse''. 
In the ``Different vs Similar'' analysis, pairs of PPC from the dataset are split into two categories as pairs with (\textit{i.e., different}) and without (\textit{i.e., similar}) statistically significant differences. For a given pair of PPC, one way ANOVA followed by Tukey's honest significance difference test~\cite{tukey1949honestsignftest} is used to measure the statistical significance of the differences. The assumption is that the absolute differences in metric predictions for ``different'' pairs should be larger than the ``similar'' pairs. Receiver Operating Characteristic (ROC) analysis is used to quantify the performance, and expressed as Area Under the ROC Curve (AUC). 
In the ``Better vs Worse'' analysis, pairs identified as ``different'' in the first stage are used. In this stage, the aim is to measure how well the metrics identify the better PPC in pairs with statistically significant difference. Metric performance in this stage can be expressed as the correct classification percentage as well as AUC values, similar to the first stage. 
An illustrative example of how metric score differences should be distributed for each stage is depicted in \figref{fig:ideal_distributions}. In ``Different vs Similar'' analysis, higher metric score differences are expected for ``Different'' pairs and lower differences for ``Similar'' pairs. In the ``Better vs Worse'' analysis, for pairs categorized as ``Better'', positive metric score differences are expected, indicating that the first point cloud in the pair is better than the second one. Conversely, for pairs categorized as ``Worse'', negative metric score differences are expected, indicating that the first point cloud in the pair is worse than the second one.

\begin{table*}[]
\caption{Columns "All" present the Pearson and Spearman correlation coefficients between the listed metric predictions and MOS (with ZREC\cite{zhu2023zrec}) of the all PPC in the \textbf{broad quality range}. Moreover, metric performances for each compression algorithm are also individually reported in the corresponding columns.}
\label{tab:metric_corr_fqrange}
\resizebox{\textwidth}{!}{%
\begin{tabular}{cl|cc|cccccccc}
\hline
\multicolumn{1}{l}{} &
   &
  \multicolumn{2}{c|}{All} &
  \multicolumn{2}{c}{GeoCNN} &
  \multicolumn{2}{c}{\begin{tabular}[c]{@{}c@{}}GPCC\\ Predlift\end{tabular}} &
  \multicolumn{2}{c}{\begin{tabular}[c]{@{}c@{}}GPCC\\ Raht\end{tabular}} &
  \multicolumn{2}{c}{VPCC} \\ \hline
\multicolumn{1}{l}{Category} &
  Metric &
  PLCC &
  SROCC &
  PLCC &
  SROCC &
  PLCC &
  SROCC &
  PLCC &
  SROCC &
  PLCC &
  SROCC \\ \hline
\multirow{14}{*}{\begin{tabular}[c]{@{}c@{}}Image \\ Based\end{tabular}} &
  MSE &
  0.2387 &
  0.2226 &
  0.4099 &
  0.4183 &
  0.3349 &
  0.2578 &
  0.3194 &
  0.2387 &
  0.1306 &
  0.0649 \\
 &
  PSNR &
  0.2488 &
  0.2269 &
  0.4220 &
  0.4231 &
  0.3395 &
  0.2621 &
  0.3238 &
  0.2429 &
  0.1149 &
  0.0691 \\
 &
  SSIM\cite{wang2004ssim} &
  0.6119 &
  0.5431 &
  0.5496 &
  0.5373 &
  0.7491 &
  0.6145 &
  0.7599 &
  0.6365 &
  0.4034 &
  0.3675 \\
 &
  MS-SSIM\cite{wang2003msssim} &
  0.5481 &
  0.4607 &
  0.4944 &
  0.4813 &
  0.6744 &
  0.5286 &
  0.6847 &
  0.5473 &
  0.3553 &
  0.2951 \\
 &
  FSIM\cite{zhang2011fsim} &
  0.6335 &
  0.5612 &
  0.5724 &
  0.5714 &
  0.7610 &
  0.6308 &
  0.7683 &
  0.6501 &
  0.4042 &
  0.3748 \\
 &
  FSIMc\cite{zhang2011fsim} &
  0.6334 &
  0.5608 &
  0.5713 &
  0.5703 &
  0.7607 &
  0.6298 &
  0.7679 &
  0.6495 &
  0.4040 &
  0.3745 \\
 &
  GMSD\cite{xue2014gmsd} &
  0.6716 &
  0.6113 &
  0.6331 &
  0.6336 &
  0.7988 &
  0.6718 &
  0.7944 &
  0.6854 &
  0.4556 &
  0.4390 \\
 &
  D-JNDQ\cite{ak2022djndq} &
  0.6732 &
  0.6313 &
  \textbf{0.7215} &
  \textbf{0.7229} &
  0.7891 &
  0.6762 &
  0.8017 &
  0.7001 &
  0.4352 &
  0.4402 \\
 &
  MW-PSNR-FR\cite{stankovic2015mwpsnrFR} &
  0.3479 &
  0.3296 &
  0.4590 &
  0.4577 &
  0.4471 &
  0.3766 &
  0.4366 &
  0.3668 &
  0.1956 &
  0.1624 \\
 &
  MW-PSNR-RR\cite{stankovic2016mwpsnrRR} &
  0.5110 &
  0.4975 &
  0.5704 &
  0.5647 &
  0.6250 &
  0.5591 &
  0.6275 &
  0.5600 &
  0.3011 &
  0.3107 \\
 &
  ADM2\cite{li2011adm2} &
  0.7283 &
  0.6520 &
  0.6434 &
  0.6216 &
  0.8408 &
  0.6887 &
  0.8362 &
  0.7073 &
  0.5513 &
  0.5386 \\
 &
  VIF-scale3\cite{sheikh2006vif} &
  0.6492 &
  0.5947 &
  0.5962 &
  0.6035 &
  0.7705 &
  0.6525 &
  0.7745 &
  0.6717 &
  0.4311 &
  0.4249 \\
 &
  VMAF\cite{netflix2018VMAF} &
  \textbf{0.7419} &
  \textbf{0.6686} &
  0.6572 &
  0.6391 &
  \textbf{0.8540} &
  \textbf{0.7091} &
  \textbf{0.8532} &
  \textbf{0.7310} &
  \textbf{0.5541} &
  \textbf{0.5412} \\
 &
  FVVDP\cite{mantiuk2021fvvdp} &
  0.6936 &
  0.6417 &
  0.6312 &
  0.6457 &
  0.8197 &
  0.6983 &
  0.8300 &
  0.7255 &
  0.4676 &
  0.4631 \\ \hline
\multirow{6}{*}{\begin{tabular}[c]{@{}c@{}}Color\\ Based\end{tabular}} &
  Color-Y-PSNR &
  0.5376 &
  0.5282 &
  0.2166 &
  0.2448 &
  0.7407 &
  0.7018 &
  0.7567 &
  0.7208 &
  0.4251 &
  0.4335 \\
 &
  Color-U-PSNR &
  0.5432 &
  0.5105 &
  0.2585 &
  0.1254 &
  0.6814 &
  0.6529 &
  0.6602 &
  0.6291 &
  0.3838 &
  0.3751 \\
 &
  Color-V-PSNR &
  0.5729 &
  0.5411 &
  0.3034 &
  0.2661 &
  0.7014 &
  0.6755 &
  0.6893 &
  0.6614 &
  0.4428 &
  0.4238 \\
 &
  PointSSIM-ColorAB\cite{evangelos2020pointssim} &
  \textbf{0.7291} &
  0.6907 &
  0.5855 &
  0.3936 &
  0.8021 &
  0.7962 &
  0.8249 &
  0.8278 &
  \textbf{0.7378} &
  \textbf{0.7675} \\
 &
  PointSSIM-ColorBA\cite{evangelos2020pointssim} &
  0.7241 &
  \textbf{0.6928} &
  \textbf{0.6044} &
  \textbf{0.4312} &
  \textbf{0.8033} &
  \textbf{0.7984} &
  \textbf{0.8257} &
  \textbf{0.8288} &
  0.7287 &
  0.7611 \\
 &
  PointSSIM-ColorSym\cite{evangelos2020pointssim} &
  0.7250 &
  0.6919 &
  0.6021 &
  0.4296 &
  0.8028 &
  0.7975 &
  0.8253 &
  0.8279 &
  0.7317 &
  0.7633 \\ \hline
\multirow{11}{*}{\begin{tabular}[c]{@{}c@{}}Geometry\\ Based\end{tabular}} &
  p2point-MSE\cite{mekuria2016point2point} &
  0.8427 &
  0.7759 &
  0.6060 &
  0.6114 &
  \textbf{0.9718} &
  0.8863 &
  \textbf{0.9707} &
  0.9002 &
  0.7221 &
  0.7154 \\
 &
  p2point-PSNR\cite{mekuria2016point2point} &
  0.6827 &
  0.4850 &
  0.2798 &
  0.2257 &
  0.7697 &
  0.5876 &
  0.7822 &
  0.6164 &
  0.5172 &
  0.4294 \\
 &
  p2plane-MSE\cite{tian2017point2plane} &
  0.8866 &
  \textbf{0.8370} &
  \textbf{0.6865} &
  \textbf{0.6415} &
  0.9681 &
  \textbf{0.8886} &
  0.9678 &
  \textbf{0.9028} &
  0.7915 &
  0.8068 \\
 &
  p2plane-PSNR\cite{tian2017point2plane} &
  0.7001 &
  0.5164 &
  0.3216 &
  0.3113 &
  0.7576 &
  0.5794 &
  0.7708 &
  0.6071 &
  0.5880 &
  0.4816 \\
 &
  pl2plane-Mean\cite{alexiou2018plane2plane} &
  0.1393 &
  0.1272 &
  0.3390 &
  0.1193 &
  0.1730 &
  0.1368 &
  0.1753 &
  0.1610 &
  0.1029 &
  0.0994 \\
 &
  pl2plane-RMS\cite{alexiou2018plane2plane} &
  0.1197 &
  0.1002 &
  0.2205 &
  0.0553 &
  0.1511 &
  0.1195 &
  0.1565 &
  0.1448 &
  0.0942 &
  0.0777 \\
 &
  pl2plane-MSE\cite{alexiou2018plane2plane} &
  0.1189 &
  0.1002 &
  0.3334 &
  0.2193 &
  0.1508 &
  0.1195 &
  0.1562 &
  0.1448 &
  0.0942 &
  0.0777 \\
 &
  PCQM\cite{meynet2020pcqm} &
  \textbf{0.8878} &
  0.8102 &
  0.4475 &
  0.2965 &
  0.9510 &
  0.8746 &
  0.9584 &
  0.8953 &
  \textbf{0.8507} &
  \textbf{0.8332} \\
 &
  PointSSIM-GeomAB\cite{evangelos2020pointssim} &
  0.7760 &
  0.7196 &
  0.5497 &
  0.5469 &
  0.9067 &
  0.8510 &
  0.9091 &
  0.8784 &
  0.5439 &
  0.5527 \\
 &
  PointSSIM-GeomBA\cite{evangelos2020pointssim} &
  0.7644 &
  0.7145 &
  0.5572 &
  0.5783 &
  0.9075 &
  0.8477 &
  0.9107 &
  0.8757 &
  0.5102 &
  0.5148 \\
 &
  PointSSIM-GeomSym\cite{evangelos2020pointssim} &
  0.7731 &
  0.7226 &
  0.5566 &
  0.5767 &
  0.9094 &
  0.8510 &
  0.9116 &
  0.8782 &
  0.5677 &
  0.5493 \\ \hline
\end{tabular}%
}
\end{table*}

\subsection{Broad Quality Range Evaluation}
\label{subsec:broad_quality_eval}

Broad quality range evaluation scenario is the generic and commonly used use-case in the literature and it typically involves calculating metric performances with the traditional correlation measures (\textit{e.g.,} PLCC, SROCC, etc.). The entire dataset is used for this evaluation, and results reported as PLCC and SROCC values between the metric predictions and the collected MOS.

Table \ref{tab:metric_corr_fqrange} presents the PLCC and SROCC of each metric in this evaluation scenario. The metrics are categorized into three group based on input type that they are operating on, as previously discussed in Section \ref{subsec:selected_metrics}. The first two column show the metrics' PLCC and SROCC scores on the entire dataset. Additionally, metric performances were evaluated for individual compression algorithms and the results are presented in subsequent columns as indicated above each column. 

PCQM~\cite{meynet2020pcqm} and p2plane-MSE~\cite{tian2017point2plane} exhibit the best performance on the entire dataset among the selected metrics, despite their poorer performance in predicting GeoCNN compression distortions. Among color-based metrics, we again notice a similar pattern on the accuracy of metrics when it comes to GeoCNN compression distortions. PointSSIM~\cite{evangelos2020pointssim} variants perform relatively better than other metrics in this category. 

Simple image-based metrics (\textit{e.g., MSE, PSNR, SSIM~\cite{wang2004ssim}, MS-SSIM~\cite{wang2003msssim}}) have low accuracy across all compression categories and consequently on the whole dataset. VMAF~\cite{netflix2018VMAF} shows the best performance among image-based metrics in the whole dataset. We also observe a general trend among image-based metrics towards a lack of accuracy on VPCC compression distortions. On another note, we observe that D-JNDQ~\cite{ak2022djndq} performs the best to predict GeoCNN distortions among the selected metrics, despite not being retrained on the dataset.

\begin{table}[]
\centering
\caption{Columns present the Pearson and Spearman correlation coefficients between the listed metric predictions and MOS (with ZREC\cite{zhu2023zrec}) of the PPC in the \textbf{high-quality range} where $MOS \geq 3.5$.}
\label{tab:metric_corr_hqrange}
\resizebox{0.83\columnwidth}{!}{%
\begin{tabular}{c|l|lc}
\hline
Category & Metric                                          & PLCC            & SROCC           \\ \hline
\multirow{14}{*}{\begin{tabular}[c]{@{}c@{}}Image \\ Based\end{tabular}}   & MSE                                      & 0.1187 & 0.0920 \\
         & PSNR                                            & 0.1476          & 0.1397          \\
         & SSIM\cite{wang2004ssim}                         & 0.2901          & 0.2119          \\
         & MS-SSIM\cite{wang2003msssim}                    & 0.2697          & 0.1600          \\
         & FSIM\cite{zhang2011fsim}                        & 0.3100          & 0.2508          \\
         & FSIMc\cite{zhang2011fsim}                       & 0.3100          & 0.2502          \\
         & GMSD\cite{xue2014gmsd}                          & 0.3503          & 0.2993          \\
         & D-JNDQ\cite{ak2022djndq}                        & \textbf{0.4120} & \textbf{0.3771} \\
         & MW-PSNR-FR\cite{stankovic2015mwpsnrFR}          & 0.1548          & 0.1490          \\
         & MW-PSNR-RR\cite{stankovic2016mwpsnrRR}          & 0.2795          & 0.2633          \\
         & ADM2\cite{li2011adm2}                           & 0.3171          & 0.2750          \\
         & VIF-scale3\cite{sheikh2006vif}                  & 0.2697          & 0.1600          \\
         & VMAF\cite{netflix2018VMAF}                      & 0.3466          & 0.3067          \\
         & FVVDP\cite{mantiuk2021fvvdp}                    & 0.3598          & 0.3278          \\ \hline
\multirow{6}{*}{\begin{tabular}[c]{@{}c@{}}Color\\ Based\end{tabular}}     & Color-Y-PSNR                             & 0.2751 & 0.2728 \\
         & Color-U-PSNR                                    & 0.1808          & 0.1808          \\
         & Color-V-PSNR                                    & 0.2192          & 0.2051          \\
         & PointSSIM-ColorAB\cite{evangelos2020pointssim}  & 0.4530          & 0.3938          \\
         & PointSSIM-ColorBA\cite{evangelos2020pointssim}  & 0.4545          & \textbf{0.4052} \\
         & PointSSIM-ColorSym\cite{evangelos2020pointssim} & \textbf{0.4547} & 0.4004          \\ \hline
\multirow{11}{*}{\begin{tabular}[c]{@{}c@{}}Geometry\\ Based\end{tabular}} & p2point-MSE\cite{mekuria2016point2point} & 0.4898 & 0.4221 \\
         & p2point-PSNR\cite{mekuria2016point2point}       & 0.1077          & 0.0774          \\
         & p2plane-MSE\cite{tian2017point2plane}           & \textbf{0.5708} & \textbf{0.5418} \\
         & p2plane-PSNR\cite{tian2017point2plane}          & 0.1067          & 0.1512          \\
         & pl2plane-Mean\cite{alexiou2018plane2plane}      & 0.0292          & 0.0511          \\
         & pl2plane-RMS\cite{alexiou2018plane2plane}       & 0.0770          & 0.0939          \\
         & pl2plane-MSE\cite{alexiou2018plane2plane}       & 0.0494          & 0.0642          \\
         & PCQM\cite{meynet2020pcqm}                       & 0.5038          & 0.4775          \\
         & PointSSIM-GeomAB\cite{evangelos2020pointssim}   & 0.4418          & 0.4527          \\
         & PointSSIM-GeomBA\cite{evangelos2020pointssim}   & 0.4466          & 0.4617          \\
         & PointSSIM-GeomSym\cite{evangelos2020pointssim}  & 0.4455          & 0.4615          \\ \hline
\end{tabular}%
}
\end{table}

\subsection{High Quality Range Evaluation}
\label{subsec:high_quality_eval}

High-quality range evaluation is crucial for applications that aim to deliver top-tier content, such as high-quality streaming and digital twins. It's important to note that metrics that perform well in the general quality range may not exhibit the same level of accuracy in the high-quality range. Therefore, we conducted an analysis to assess the accuracy of quality metrics specifically on the high-quality part of the dataset, where the MOS is greater than or equal to $3.5$. This evaluation helps identify which metrics excel in scenarios where maintaining exceptionally high quality is a priority.

In this evaluation scenario, metric performances are relatively low overall, as indicated in Table \ref{tab:metric_corr_hqrange}. However, the relative order of the metrics in terms of their performances remains relatively consistent with the broad quality range evaluation. p2plane-MSE~\cite{tian2017point2plane} performs the best overall while D-JNDQ~\cite{ak2022djndq} is the best performing image-based metric and the PointSSIM-ColorBA\cite{evangelos2020pointssim} is the best performing color-based metric for this use-case. 

Various factors contribute to the low performance of metrics on the high quality range. First of all, most metrics are developed to account for a wider range of distortion intensities. Features, parameters, and modules used in a wide quality range are not as related to a more narrow range. A narrow quality range results in lack of statistically significant difference between the data points, resulting in a more difficult problem.

\begin{figure*}[!t]
    \centering
    \includegraphics[width=\textwidth]{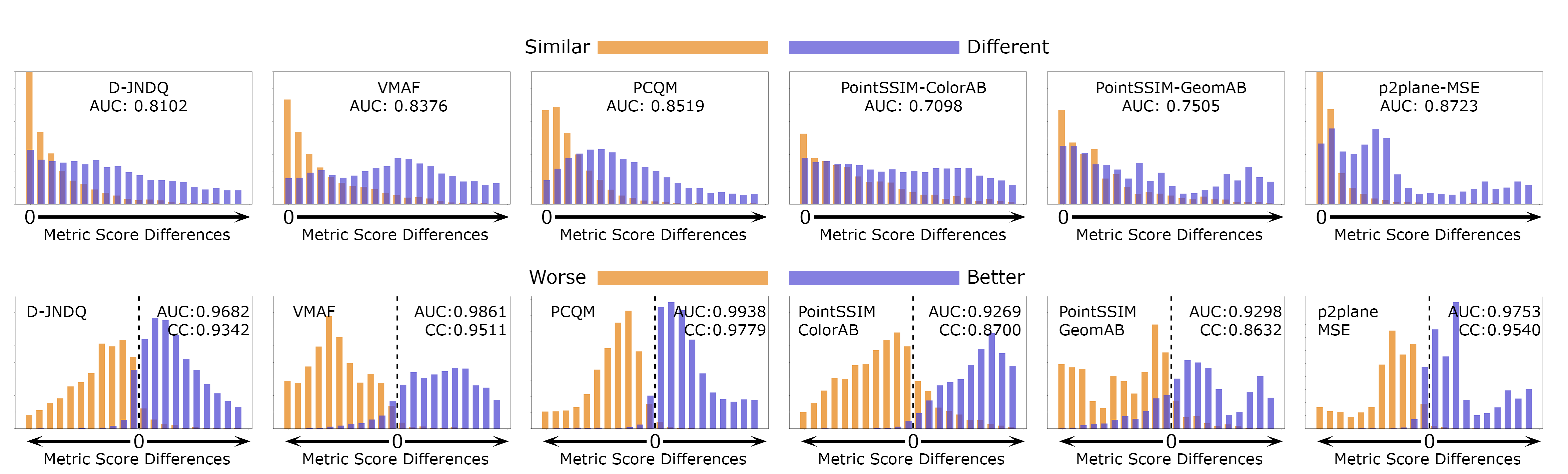}
    \caption{The plots in the top row depict the metric score differences for pairs categorized as ``Different'' and ``Similar'' for the \textbf{Intra-SRC evaluation}. Each metric's score differences are individually normalized within minimum and maximum ranges. The height of the bars represents the number of occurrences, and each bar ranges between 0 and 1500. The metric names are indicated at the top of each plot, and the AUC values are reported below each metric name. Similarly, the plots in the bottom row show the metric score differences for pairs categorized as ``Better'' and ``Worse'' for the Intra-SRC evaluation. Again, the metric score differences are individually normalized within minimum and maximum ranges. The height of the bars denotes the number of occurrences, and each bar ranges between 0 and 800. The metric names are indicated at the top left corner of each plot and the AUC and CC are reported in the top right corner of each plot.}
    \label{fig:krasula_eval_results}
\end{figure*}

\subsection{Intra-SRC Evaluation}
\label{subsec:intra_src_eval}

Intra-SRC evaluation focuses on assessing metrics' performance when comparing PPC derived from a single pristine SRC at various compression levels. It allows us to gauge how well metrics can discriminate between different compression levels originating from the same SRC, helping us optimize processes that rely on such discrimination. This evaluation scenario is valuable for applications like fine-tuning compression and enhancement algorithms, training machine learning models for end-to-end processing, and other situations where fidelity is a primary concern.

Prior to the analysis, we preprocess the subjective scores as described in Krasula's method\cite{krasula2016metriceval}. First, 20 PPCs were paired within each source point cloud, generating $(20\times(20-1)/2)$ pairs per SRC. In total, we end up with 14143 pairs. Afterwards, a one-way ANOVA test is applied to individual scores collected for each stimulus in each pair, followed by Tukey's Honest test. 5019 pairs among the total 14143 were identified as ``Similar'' whereas 9124 contains a statistically significant different between the two PPC and thus identified as ``Different''. From those ``Different'' pairs, we split them into two roughly equal-sized groups as ``Better'' and ``Worse'' depending on the order of the pair. There are 4075 ``Better'' and 5049 ``Worse'' pairs. 

Due to space limitations of the manuscript, we report the result of the analysis only on the selected 6 metrics among the initial list presented in Section \ref{subsec:selected_metrics}. The rest of the results can be acquired via the provided scripts in the GitHub repository of the dataset.

\textbf{Different vs Similar Analysis:}
The top row in \figref{fig:krasula_eval_results} presents the results of the analysis as histograms of metric score differences for ``Different'' and ``Similar'' pairs. We expect better-performing metrics to provide metric score distributions similar to the ideal case as depicted in \figref{fig:ideal_distributions}. Additionally, performance of each metric quantified with AUC values, reported under each metric name. We observe a better performance from PCQM, providing a higher AUC value and a very similar distribution to the ideal case. Statistical significance tests on this task also reveals that PCQM performs significantly better than all other metrics in ``Different vs Similar'' task.

\textbf{Better vs Worse Analysis:} Similar to the previous stage, bottom row of the \figref{fig:krasula_eval_results} presents the results as histograms of metric score differences and quantifies the performance of each metric with AUC and CC values. We observe that most metrics perform relatively well on identifying ``Better'' and ``Worse'' pairs apart. PCQM shows a very similar distribution to the ideal case depicted in \figref{fig:ideal_distributions} as also reflected by the AUC and CC values.

\section{Conclusion}
\label{sec:conclusion}

We conducted a large-scale crowdsourcing study on point cloud compression quality assessment. To the best of our knowledge, this is the largest publicly available point cloud quality assessment dataset. It contains 75 source point clouds, each compressed with 4 different compression algorithms resulting in nearly 1500 processed point clouds. More than 3500 naive observers participated to the experiment. 

Our study revealed several noteworthy insights regarding objective quality metrics' performances. While most point cloud objective quality metrics perform well in predicting GPCC distortions, the majority of the metrics still struggle with VPCC distortions. Furthermore, an even larger majority fall short in assessing the quality GeoCNN distortions, a learning-based compression algorithm. This highlights a pressing need for improved quality metrics capable of accurately evaluating learning-based compression distortions. Given the scarcity of learning-based compression algorithms in publicly available datasets (see Table \ref{tab:DatasetOverview}), the need for better quality metrics that can accurately predict learning-based compression distortions is revealed in this work. 

Additionally, we observed significant room for improvement in metrics designed for high-quality content. The correlations between the MOS and the best-performing metric predictions (p2plane-MSE in this scenario) remained below 0.60. On a more positive note, our intra-SRC evaluation scenario yielded more promising results. The best-performing metrics demonstrated more promising results for distinguishing the higher-quality point cloud. Nevertheless, there remains room for improvement in predicting the statistical significance of quality differences between point clouds.

As part of our commitment to advancing point cloud quality assessment, we are making our dataset publicly available. This dataset includes mean and individual opinion scores, along with scripts for metric evaluation in various scenarios. It also comprises all point clouds and their associated video renderings, forming what we call the BASICS dataset. We believe that the release of the BASICS dataset will contribute significantly to the improvement of existing point cloud quality metrics, the development of more robust ones, and the resolution of the challenges highlighted in this study. Finally, we plan to address some of the limitations in future studies by including dynamic PCs, a wider variety of semantic categories and styles, and other learning-based compression algorithms which performs both geometry and attribute coding.


\section*{Acknowledgments}

This work has received funding from the European Union's Horizon 2020 research and innovation programme under the Marie Sklodowska-Curie Grant Agreement No. 765911 (RealVision) and from Science Foundation Ireland (SFI) under the Grant Number 15/RP/27760.

\bibliographystyle{IEEEtran}
\bibliography{refs}

\begin{IEEEbiography}[{\includegraphics[width=1in,height=1.25in,clip,keepaspectratio]{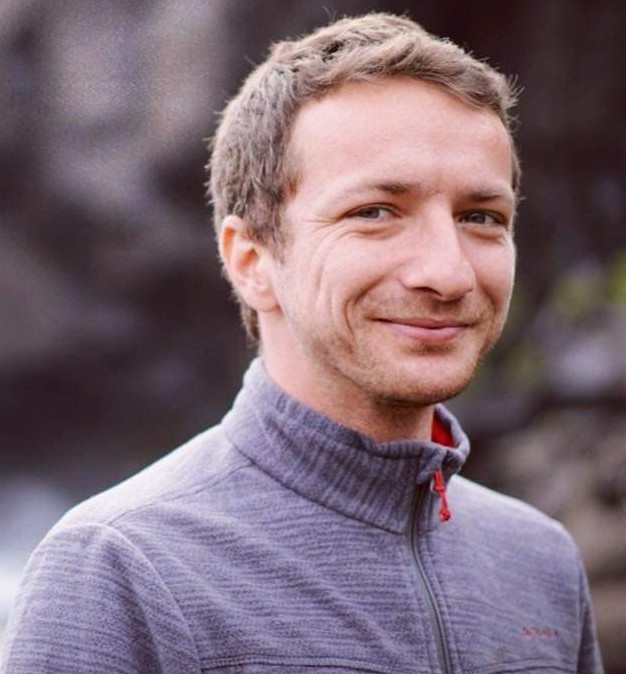}}]{Ali Ak} (Member, IEEE) is a Postdoctoral Researcher at Nantes Universit\'e, Nantes, France. His current interests are subjective and objective evaluation of visual content such as videos, point clouds, volumetric videos, and light fields. He regularly contributes to standardization activities at IEEE and VQEG. He completed his PhD at Nantes Universit\'e focusing on perceptual evaluation on immersive multimedia content. He received his M.Sc. degree from Visual Computing master program at Nantes Universit\'e. He completed his B.Sc. studies on Mathematics at Bilkent University, Ankara, Turkey. 
\end{IEEEbiography}

\begin{IEEEbiography}[{\includegraphics[width=1in,height=1.25in,clip,keepaspectratio]{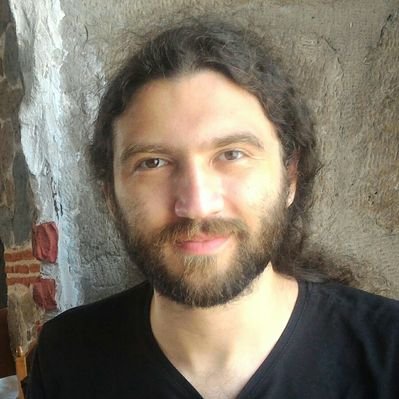}}]{Emin Zerman} is an Associate Senior Lecturer at Mid Sweden University. Previously, he has been working as a postdoctoral research fellow at Trinity College Dublin, Ireland, working in V-SENSE project on immersive imaging technologies. He received his Ph.D. degree (2018) in Signals and Images from Télécom ParisTech, France, and his M.Sc. degree (2013) and B.Sc. degree (2011) in Electrical and Electronics Engineering from the Middle East Technical University, Turkey. He is interested in multimedia quality assessment, human visual perception, data visualization, user interaction, immersive multimedia, and 3D technologies.
\end{IEEEbiography}

\begin{IEEEbiography}[{\includegraphics[width=1in,height=1.25in,clip,keepaspectratio]{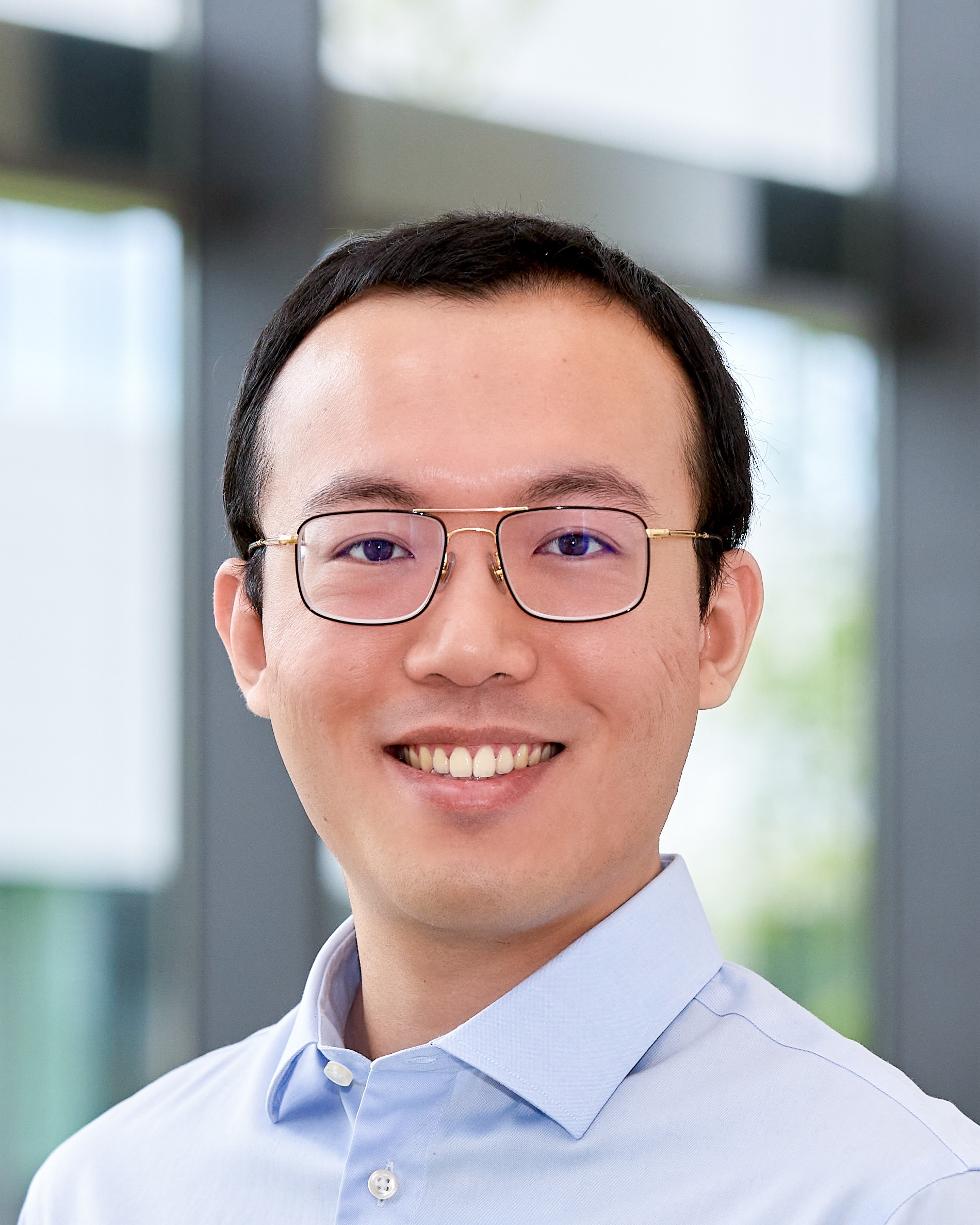}}]{Maurice Quach}  is a Research Engineer at the Bosch Center for Artificial Intelligence, Renningen, Germany. He completed his Ph.D. (2022) on Deep learning-based Point Cloud Compression at the University of Paris-Saclay, France. He received his M.Sc. degree (2018) in Computer Science at the University of Technology of Compi\'egne, France.
\end{IEEEbiography}

\begin{IEEEbiography}[{\includegraphics[width=1in,height=1.25in,clip,keepaspectratio]{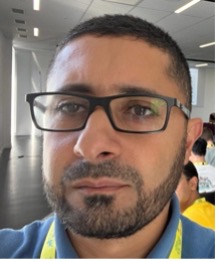}}] {Aladine Chetouani} (Senior Member, IEEE) received his master’s degree in computer science from the University Pierre and Marie Curie, France, in 2005, and his Ph.D. degree in image processing from the University of Paris 13, France, in 2010. From 2010 to 2011, he was a postdoctoral researcher with the L2TI Laboratory, Paris 13 University. In 2020, he benefited from a CNRS delegation year at the L2S laboratory, Centrale Supélec, Université Paris Saclay, France. He is currently an associate professor with the Laboratory PRISME, Orleans, France. He also received the Habilitation degree entitled “On the use of visual attention and deep learning for blind quality assessment of multimedia contents” from Université d'Orléans in 2020. He is an IEEE senior member. He is co-author of more than 120 research publications in international refereed journals and conference proceedings. He is a member of the MMSP and IVMSP technical committees of the IEEE Signal Processing Society. He is an associate editor of IEEE Transactions on Multimedia. He was the general chair of Content-Based Multimedia Indexing 2023 conference (CBMI 2023). He is a member of the GDR-IASIS committee (comité de direction). His present interests are in image quality, perceptual analysis, visual attention using deep learning models for different multimedia content (image: real, medical, painting, etc.; stereo, 3D: meshes and point clouds).
\end{IEEEbiography}

\begin{IEEEbiography}[{\includegraphics[width=1in,height=1.25in,clip,keepaspectratio]{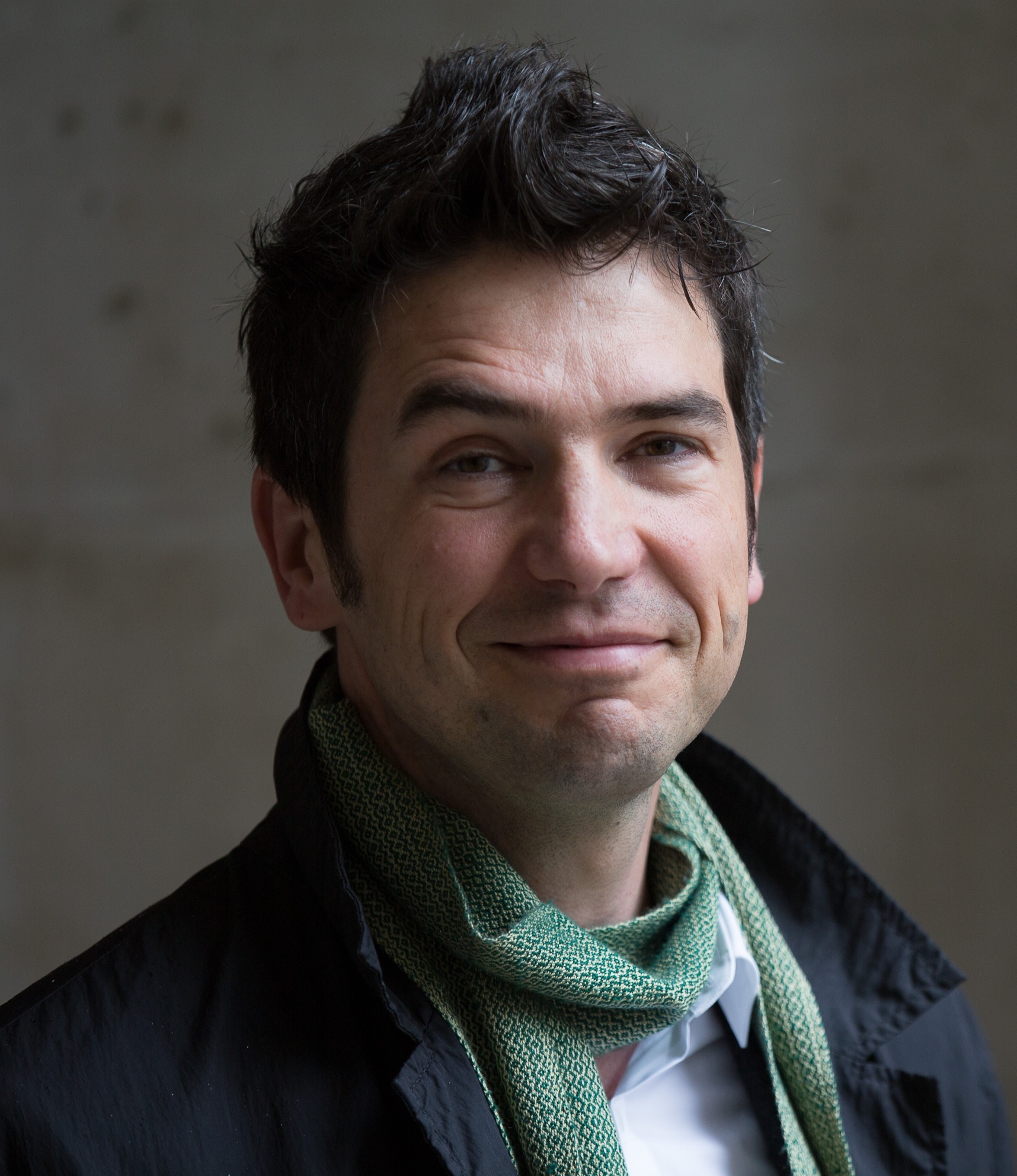}}] {Aljosa Smolic} is Professor in the Computer Science Department of Hochschule Luzern in Switzerland and Co-Head of the Immersive Realities Research Lab. Before he was Professor of Creative Technologies at Trinity College Dublin heading the research group V-SENSE, Senior Research Scientist and Group Leader at Disney Research Zurich, and Scientific Project Manager and Group Leader at Fraunhofer HHI. He is also a Co-Founder of the company Volograms, which commercializes volumetric video technology. Prof. Smolic’s expertise is in the broad area of visual computing (covering image/video processing, computer vision, computer graphics) with a focus on immersive XR technologies. He published 250+ scientific papers and book chapters, holds 35+ patents and received several awards and recognitions for his research, including the IEEE ICME Star Innovator Award 2020 for his contributions to volumetric video content creation. Prof. Smolic served as Associate Editor of the IEEE Transactions on Image Processing and the Signal Processing: Image Communication journal. He was Guest Editor for the Proceedings of the IEEE, IEEE Transactions on CSVT, IEEE Signal Processing Magazine, and other scientific journals.
\end{IEEEbiography}

\begin{IEEEbiography}[{\includegraphics[width=1in,height=1.25in,clip,keepaspectratio]{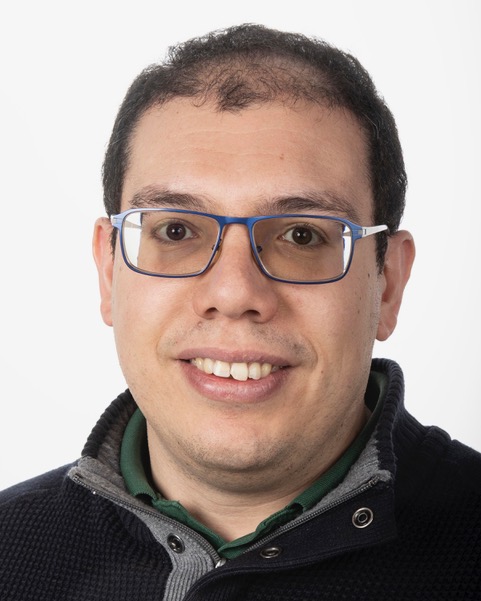}}] {Giuseppe Valenzise} (Senior Member, IEEE) is a CNRS researcher and head of the Multimedia and Networking team at the Laboratoire des Signaux et Systèmes of CentraleSupelec and Université Paris-Saclay. He got his PhD from Politecnico di Milano, Italy, in 2011. He was with Telecom Paristech from 2012 to 2016. His research interests span different fields of image and video processing, including traditional and learning-based image and video compression, immersive video and visual quality assessment. He is co-author of more than 100 research publications in these domains. He is the recipient of the EURASIP Early Career Award 2018. Giuseppe serves/has served as Associate Editor for IEEE Transactions on Circuits and Systems for Video Technology, IEEE Transactions on Image Processing, Elsevier Signal Processing: Image communication. He is the Chair of the MMSP technical committee of the IEEE Signal Processing Society and a previous member of the SPS IVMSP technical committee and of the Technical Area Committee on Visual Information Processing of EURASIP.
\end{IEEEbiography}

\begin{IEEEbiography}[{\includegraphics[width=1in,height=1.25in,clip,keepaspectratio]{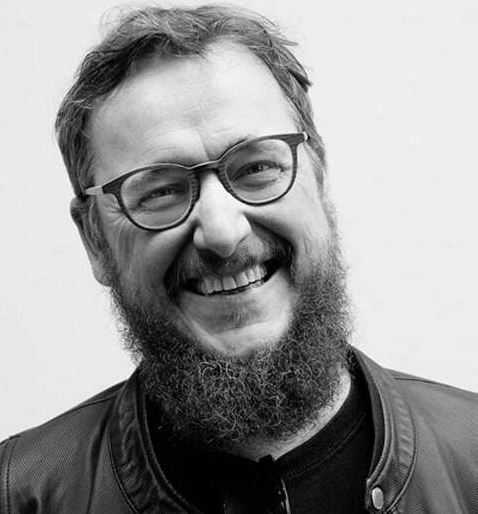}}]{Patrick Le Callet} (Fellow, IEEE) is Full Professor at Polytech Nantes / Université de Nantes (Engineering School) in EE and CS. He is also senior member of the Institut Universitaire de France (IUF). He is serving in the Steering Board of the CNRS LS2N lab (450 researchers). From 2015 to 2022,  he was the scientific director of the cluster “Ouest Industries Créatives” gathering more than 10 institutions (including 3 universities). “Ouest Industries Créatives” aims to strengthen Research, Education and Innovation of the Region Pays de Loire in the field of Experience. He is leading a muti disciplinary team to conduct inter disciplinary dealing with the application of human perception in media processing and cognitive computing, including all form of artificial intelligence. His current centers of interest are focused on twin transition addressing Quality of Experience assessment for sustainable visual communication, Quality of Life measurement for better healthcare or inclusive technologies. He is co-author of more than 350 publications and communications and co-inventor of 16 international patents on these topics. He serves or has been served as associate editor or guest editor for several Journals such as IEEE Signal Processing Magazine, IEEE TIP, IEEE STSP, IEEE TCSVT, SPRINGER EURASIP Journal on Image and Video Processing, and SPIE JEI. He has served in IEEE IVMSP-TC (2015- to present) and IEEE MMSP-TC (2015-to present) and the chair of EURASIP TAC (Technical Area Team) on Visual Image Processing. He is chairing activities in Standard (VQEG and IEEE-SA). He is co-recipient of an Emmy Award in 2020 for his work on development of Perceptual metrics for video encoding optimization.
\end{IEEEbiography}

\vfill

\begin{table*}[htbp]
\caption{Appendix Table: List of all selected PCs with several attributes, including what the content is, the creator, and copyright license. There were three different categories for point clouds: humans \& animals (\textit{HA}), inanimate objects (\textit{IO}), and buildings \& landscape (\textit{BL}). Format column indicates if the 3D model was collected in point cloud (\textit{P}) or 3D mesh format (\textit{M}), followed by the file extension (e.g., \textit{M-obj}). Capture column indicates if the model was scanned from real world or modeled virtually. The total number of points in reference PCs are indicated in Pt\# column.}
\label{tab:listAppendix}
\centering
\scriptsize
\begin{tabular}{ccccclll} \hline \hline
PC\# & Cat & Format & Capture & Pt\#    & Contents              & Creator                                                                                                                                                                 & Copyright                 \\ \hline
p01 & HA  & P-ply  & Scan    & 85.6k  & People                & XD Productions                                                                                                                                                          & CC BY-NC \\
p02 & HA  & P-ply  & Scan    & 525.3k & People                & XD Productions                                                                                                                                                          & CC BY-NC \\
p03 & HA  & P-ply  & Scan    & 334.0k & People                & XD Productions                                                                                                                                                          & CC BY-NC \\
p04 & HA  & P-ply  & Scan    & 434.1k & People                & XD Productions                                                                                                                                                          & CC BY-NC \\
p05 & HA  & P-ply  & Scan    & 151.8k & People                & XD Productions                                                                                                                                                          & CC BY-NC \\
p06 & HA  & P-ply  & Scan    & 168.0k & People                & XD Productions                                                                                                                                                          & CC BY-NC \\
p07 & HA  & P-ply  & Scan    & 182.3k & People                & XD Productions                                                                                                                                                          & CC BY-NC \\
p08 & HA  & P-ply  & Scan    & 150.0k & People                & XD Productions                                                                                                                                                          & CC BY-NC \\
p09 & HA  & P-ply  & Scan    & 509.0k & People                & XD Productions                                                                                                                                                          & CC BY-NC \\
p10 & HA  & P-ply  & Scan    & 359.4k & People                & XD Productions                                                                                                                                                          & CC BY-NC \\
p11 & HA  & M-obj  & Scan    & 1.14M  & People                & V-SENSE, TCD                                                                                                                                                            & CC BY                     \\
p12 & HA  & M-obj  & Scan    & 880.6k & People                & V-SENSE, TCD                                                                                                                                                            & CC BY                     \\
p13 & HA  & M-obj  & Scan    & 1.33M  & People                & V-SENSE, TCD                                                                                                                                                            & CC BY                     \\
p14 & HA  & M-obj  & Scan    & 977.5k & People                & V-SENSE, TCD                                                                                                                                                            & CC BY                     \\
p15 & HA  & M-obj  & Scan    & 901.6k & People                & V-SENSE, TCD                                                                                                                                                            & CC BY                     \\
p16 & HA  & M-obj  & Scan    & 840.7k & People                & V-SENSE, TCD                                                                                                                                                            & CC BY                     \\
p17 & HA  & M-obj  & Scan    & 970.7k & People                & V-SENSE, TCD                                                                                                                                                            & CC BY                     \\
p18 & HA  & M-obj  & Scan    & 1.00M  & People                & V-SENSE, TCD                                                                                                                                                            & CC BY                     \\
p19 & HA  & M-fbx  & Scan    & 1.80M  & Sea Turtle            & DigitalLife3D - [\href{https://sketchfab.com/3d-models/model-47a-loggerhead-sea-turtle-c438e81e796d41d9a6ae4cc147ef8d4f}{Sketchfab Link}]                               & CC BY-NC                  \\
p20 & HA  & M-fbx  & Model   & 557.4k & Dinosaur              & Joel Anderson - [\href{https://sketchfab.com/3d-models/tyrannosarus-rex-free-model-e18c433cdd1c49f8ac152348b7384037}{Sketchfab Link}]                                   & CC BY-NC                  \\
p21 & HA  & M-obj  & Scan    & 1.43M  & Elephant              & Abby Gancz - [\href{https://sketchfab.com/3d-models/elephant-in-the-rotunda-26ee59c981964681bf9f4e5eae2a3a26}{Sketchfab Link}]                                          & CC BY                     \\
p22 & HA  & M-obj  & Scan    & 741.5k & Komodo Dragon         & Araon - [\href{https://sketchfab.com/3d-models/komodo-dragon-weta-cave-c606f322fc54498ca0053331f78c17b4}{Sketchfab link}]                                               & CC BY                     \\
p23 & HA  & M-obj  & Scan    & 1.90M  & Bird                  & Virtual Museums of Małopolska - [\href{https://sketchfab.com/3d-models/ruff-8083117638b34ea9b49064b6d624a6e0}{Sketchfab link}]                                          & CC 0                      \\
p24 & HA  & M-obj  & Scan    & 1.47M  & Bird                  & Virtual Museums of Małopolska - [\href{https://sketchfab.com/3d-models/the-white-throated-dipper-6b36ba4fba994bb59efeddf09034920c}{Sketchfab link}]                     & CC 0                      \\
p25 & HA  & M-obj  & Scan    & 1.49M  & Bird                  & Virtual Museums of Małopolska - [\href{https://sketchfab.com/3d-models/lesser-grey-shrike-d63a651c8ee34bf88a80e8b11fb84ca1}{Sketchfab link}]                            & CC 0                      \\
p26 & IO  & M-obj  & Scan    & 3.51M  & Camera Bag            & Andrea Spognetta - [\href{https://sketchfab.com/3d-models/old-camera-bag-rawscan-788f8b75874f417ebde498ffd231410c}{Sketchfab link}]                                     & CC BY                     \\
p27 & IO  & M-fbx  & Model   & 754.4k & Skateboard            & Kaye Simonson - [\href{https://sketchfab.com/3d-models/skatie-ae3181c81cf34876b187b353291a2f96}{Sketchfab link}]                                                        & CC BY-NC                  \\
p28 & IO  & M-obj  & Scan    & 2.91M  & Hat                   & Andrea Spognetta - [\href{https://sketchfab.com/3d-models/hat-01-rawscan-bbed960e1cfd4ede8d0cf4755532865b}{Sketchfab link}]                                             & CC BY                     \\
p29 & IO  & M-fbx  & Model   & 1.03M  & Radio                 & Andrzej Borkowski - [\href{https://sketchfab.com/3d-models/handheld-portable-radio-walkie-talkie-282d16b8d5a84343953eb6d8d135722a}{Sketchfab link}]                     & CC BY-NC                  \\
p30 & IO  & M-fbx  & Model   & 5.03M  & Audio Recorder        & Artem P - [\href{https://sketchfab.com/3d-models/sony-tc-d5-recorder-aa2e82ca23cc415897fe66c9a3f5a3ba}{Sketchfab link}]                                                 & CC BY                     \\
p31 & IO  & M-obj  & Scan    & 3.19M  & Bread                 & Andrea Spognetta - [\href{https://sketchfab.com/3d-models/homemade-bread-rawscan-f95a2e18d8454902a779360306d680c5}{Sketchfab link}]                                     & CC BY                     \\
p32 & IO  & M-fbx  & Model   & 2.53M  & Headphone             & Halil Kantarci - [\href{https://sketchfab.com/3d-models/headphone-with-stand-4ffedc9bffad4a549f6e0a46b0f92b05}{Sketchfab link}]                                         & CC BY                     \\
p33 & IO  & M-obj  & Scan    & 612.2k & Violin                & Virtual Museums of Małopolska - [\href{https://sketchfab.com/3d-models/violin-a784af0713a643b19ffcf65194bc0fbf}{Sketchfab link}]                                        & CC 0                      \\
p34 & IO  & M-obj  & Scan    & 3.54M  & Pizza                 & Rigsters - [\href{https://sketchfab.com/3d-models/pizza-40d50989fec1460f8838b608d999ccd0}{Sketchfab link}]                                                              & CC BY                     \\
p35 & IO  & M-fbx  & Model   & 2.67M  & Shoes                 & kenprol - [\href{https://sketchfab.com/3d-models/leather-shoes-c2a5343bf18949fc8d6815034ec7f87d}{Sketchfab link}]                                                       & CC BY-NC-SA               \\
p36 & IO  & P-ply  & Scan    & 645.0k & Cactus                & GSXNet - [\href{https://sketchfab.com/3d-models/fishhook-barrel-cactus-plantpointschallenge-1f088de099d24cea9d8bba2a3d067403}{Sketchfab}]                               & CC BY                     \\
p37 & IO  & M-fbx  & Model   & 3.47M  & Derby Car             & FelikinRuslan - [\href{https://sketchfab.com/3d-models/derby-car-free-30331d8e1355492f8be6ec7bb5b556bc}{Sketchfab link}]                                                & CC BY                     \\
p38 & IO  & M-fbx  & Scan    & 4.17M  & Wagon                 & Virtual Museums of Małopolska - [\href{https://sketchfab.com/3d-models/gypsy-wagon-4-1aa5a45694b349cc860ad026c4d039f1}{Sketchfab link}]                                 & CC 0                      \\
p39 & IO  & M-obj  & Model   & 5.00M  & Lada Niva             & \textit{Unknown} - [\href{https://sketchfab.com/3d-models/lada-niva-car-low-poly-free-85a39a52751b42ca9d5482838de93d20}{Sketchfab link}]                                & CC BY-NC                  \\
p40 & IO  & M-obj  & Scan    & 2.26M  & Statue                & noe-3d.at - [\href{https://sketchfab.com/3d-models/kriegerdenkmal-b43a30c8b577434ea52bfb187164d618}{Sketchfab link}]                                                    & CC BY-NC                  \\
p41 & IO  & M-obj  & Scan    & 3.48M  & Statue                & Geoffrey Marchal - [\href{https://sketchfab.com/3d-models/baptismal-angel-kneeling-f45f01c63e514d3bad846e82af640f33}{Sketchfab link}]                                   & CC BY-NC                  \\
p42 & IO  & M-obj  & Scan    & 4.58M  & Vase                  & Virtual Museums of Małopolska - [\href{https://sketchfab.com/3d-models/hydria-apothecary-vase-7d6938c0c0b54b06a0210a982a73023e}{Sketchfab link}]                        & CC 0                      \\
p43 & IO  & M-obj  & Scan    & 5.27M  & Antique               & Virtual Museums of Małopolska - [\href{https://sketchfab.com/3d-models/casket-with-perfume-bottles-f4fe51018fa5400eb28d3d704c2d0d51}{Sketchfab link}]                   & CC 0                      \\
p44 & IO  & M-obj  & Scan    & 1.99M  & Antique               & Virtual Museums of Małopolska - [\href{https://sketchfab.com/3d-models/horn-of-salt-diggers-brotherhood-of-wieliczka-8224ff719c34483e9effed9aa0d8590e}{Sketchfab link}] & CC 0                      \\
p45 & IO  & M-fbx  & Scan    & 4.97M  & Vintage Cash Register & Virtual Museums of Małopolska - [\href{https://sketchfab.com/3d-models/cash-register-with-a-counting-machine-6b6ecc33993d49939bd4c7633c795577}{Sketchfab link}]         & CC 0                      \\
p46 & IO  & M-obj  & Scan    & 713.2k & Optical spectrometer  & Virtual Museums of Małopolska - [\href{https://sketchfab.com/3d-models/an-optical-spectrometer-1cb673f9f1044907a6261e4bb8a63464}{Sketchfab link}]                       & CC 0                      \\
p47 & IO  & M-obj  & Scan    & 4.54M  & Globe                 & Virtual Museums of Małopolska - [\href{https://sketchfab.com/3d-models/celestial-globe-341fa8a777e94883841409438756f747}{Sketchfab link}]                               & CC 0                      \\
p48 & IO  & M-obj  & Scan    & 755.5k & Wooden bike           & Virtual Museums of Małopolska - [\href{https://sketchfab.com/3d-models/wooden-bicycle-3d7fc7394cde43298de89a28c0afbaff}{Sketchfab link}]                                & CC 0                      \\
p49 & IO  & M-obj  & Scan    & 3.04M  & Fountain              & artfletch - [\href{https://sketchfab.com/3d-models/drinking-fountain-clapham-common-a362031a32fd4c5790febf2ca40da32d}{Sketchfab link}]                                  & CC BY                     \\
p50 & IO  & M-obj  & Scan    & 1.29M  & Statue                & Geoffrey Marchal - [\href{https://sketchfab.com/3d-models/tiger-devouring-a-gavial-020027b89fda46588f42ed9ff384950d}{Sketchfab link}]                                   & CC BY-NC                  \\
p51 & BL  & M-obj  & Scan    & 1.63M  & Windmill              & Lassi Kaukonen  - [\href{https://sketchfab.com/3d-models/windmill-in-halikko-kreivinmaki-finland-2a43c8bd492d4d11ad73e1d3f7ccb58a}{Sketchfab link}]                     & CC BY                     \\
p52 & BL  & M-obj  & Scan    & 4.17M  & Small Chapel          & noe-3d.at - [\href{https://sketchfab.com/3d-models/kapelle-47027db9cda04e6dac356ed171928b36}{Sketchfab link}]                                                           & CC BY-NC                  \\
p53 & BL  & M-obj  & Scan    & 2.11M  & Church                & Panomedia - [\href{https://sketchfab.com/3d-models/byzantine-church-agios-petros-c-12th-13th-7cf515073def45debaf6274a3276b39b}{Sketchfab link}]                         & CC BY                     \\
p54 & BL  & M-obj  & Scan    & 2.46M  & Church                & Global Digital Heritage - [\href{https://sketchfab.com/3d-models/santa-maria-de-melque-toledo-spain-36cacabf9f364f8cbd45ff659a6144d4}{Sketchfab link}]                  & CC BY-NC                  \\
p55 & BL  & M-obj  & Model   & 4.20M  & House                 & Renafox - [\href{https://sketchfab.com/3d-models/house-test-8ccdacecef714a4bb1e7eaa7075695c7#download}{Sketchfab link}]                                                 & CC BY-NC                  \\
p56 & BL  & P-ply  & Scan    & 2.42M  & Cabin                 & Epic\_Tree\_Store - [\href{https://sketchfab.com/3d-models/little-cabin-3b9fbbd468e54227ad7ef7e74f8f479c}{Sketchfab link}]                                                & CC BY                     \\
p57 & BL  & M-obj  & Scan    & 3.47M  & Cathedral             & Agisoft - [\href{https://sketchfab.com/3d-models/mexico-city-metropolitan-cathedral-ec98bc3741d54810a1ebf35d26c1de54}{Sketchfab link}]                                  & CC BY-NC-SA               \\
p58 & BL  & M-obj  & Scan    & 2.23M  & Museum                & Miguel Bandera - [\href{https://sketchfab.com/3d-models/palace-of-fine-arts-photogrammetry-aerial-5eb98df6ad204542af3054071e3c4883}{Sketchfab link}]                    & CC BY-NC-SA               \\
p59 & BL  & M-obj  & Scan    & 4.18M  & Building              & artfletch - [\href{https://sketchfab.com/3d-models/betsey-trotwood-pub-e9c1824a218d4befaa6acef9a2a64ce0}{Sketchfab link}]                                               & CC BY                     \\
p60 & BL  & M-obj  & Scan    & 4.27M  & Building              & artfletch - [\href{https://sketchfab.com/3d-models/euston-tap-305a234221bf426ebfbb52858879d179}{Sketchfab link}]                                                        & CC BY                     \\
p61 & BL  & M-obj  & Scan    & 3.09M  & Building              & artfletch - [\href{https://sketchfab.com/3d-models/coach-and-horses-pub-bruton-street-7310be59a72d4923b18ed7cde03ce843}{Sketchfab link}]                                & CC BY                     \\
p62 & BL  & M-obj  & Scan    & 1.67M  & Building              & UoM Digitisation Centre - [\href{https://sketchfab.com/3d-models/university-of-melbourne-gatehouse-a3af4f454dd6472b86a52b2507b7f26d}{Sketchfab link}]                   & CC BY-NC-SA               \\
p63 & BL  & M-obj  & Scan    & 3.34M  & Historic house        & Mario\_Wallner - [\href{https://sketchfab.com/3d-models/germanic-house-elsarn-ae2b6de1cc9c40c286c333aed6b2b212}{Sketchfab link}]                                         & CC BY-NC                  \\
p64 & BL  & M-obj  & Scan    & 4.21M  & Wood cabin            & Mario\_Wallner - [\href{https://sketchfab.com/3d-models/schwarzenbach-houses-with-interior-4ec009d47f184898a1ab13cc08f3a7f8}{Sketchfab link}]                            & CC BY-NC-SA               \\
p65 & BL  & M-obj  & Scan    & 1.98M  & Ancient ruin          & fuzzelhjb - [\href{https://sketchfab.com/3d-models/ancient-corinth-peirene-fountain-d7a85f6096584f02817ff255c16608e0}{Sketchfab link}]                                  & CC BY                     \\
p66 & BL  & M-obj  & Scan    & 2.58M  & Bridge                & Lassi Kaukonen  - [\href{https://sketchfab.com/3d-models/jaaninoja-bridge-in-turku-kurala-finland-65a60cc3d66249a0a33beeb6a1ad2540}{Sketchfab link}]                    & CC BY                     \\
p67 & BL  & M-obj  & Scan    & 1.93M  & Bunker                & Azad Balabanian - [\href{https://sketchfab.com/3d-models/devils-slide-bunker-pacifica-ca-52982f5a24834be48dab4057d2df1c3d}{Sketchfab link}]                             & CC BY                     \\
p68 & BL  & M-fbx  & Scan    & 1.89M  & Town landscape        & Azad Balabanian - [\href{https://sketchfab.com/3d-models/eze-village-france-a18a1449661643e6ab98640a3e7eafc1}{Sketchfab link}]                                          & CC BY                     \\
p69 & BL  & M-obj  & Scan    & 1.12M  & Stonehenge            & Arch\'eomatique - [\href{https://sketchfab.com/3d-models/stonehenge-angleterre-4824d3a2f11d41858f4e388486b36bcf}{Sketchfab link}]                                       & CC BY-NC                  \\
p70 & BL  & M-obj  & Scan    & 1.85M  & Castle                & Andrea Spognetta - [\href{https://sketchfab.com/3d-models/rocca-calascio-rawscan-7fbfe67ec4864d439f25303b50697189}{Sketchfab link}]                                     & CC BY-NC                  \\
p71 & BL  & M-obj  & Scan    & 1.80M  & Town landscape        & omnidirectional - [\href{https://sketchfab.com/3d-models/monteriggioni-high-quality-628f90b70c5547a5852862265c9b1f34}{Sketchfab link}]                                  & CC BY                     \\
p72 & BL  & M-obj  & Scan    & 1.83M  & Chateau               & Andrea Spognetta - [\href{https://sketchfab.com/3d-models/chateau-de-la-bretesche-rawscan-2eeeba7fa4824666b485bbc8e919da1c}{Sketchfab link}]                            & CC BY-NC                  \\
p73 & BL  & M-obj  & Scan    & 2.33M  & Ruins                 & Global Digital Heritage - [\href{https://sketchfab.com/3d-models/roman-temple-of-evora-935f17a3824d49f7b2505a0686450d51}{Sketchfab link}]                               & CC BY                     \\
p74 & BL  & M-fbx  & Scan    & 1.82M  & Buildings             & Andrea Spognetta - [\href{https://sketchfab.com/3d-models/skyscrapers-rawscan-158a568d06c14e388bfa34a911559c73}{Sketchfab link}]                                        & CC BY-NC                  \\
p75 & BL  & M-fbx  & Scan    & 1.20M  & Buildings             & Andrea Spognetta - [\href{https://sketchfab.com/3d-models/construction-site-rawscan-963a8cd7da8b4761ae6d0b3ca842e027}{Sketchfab link}]                                  & CC BY-NC                 \\ \hline \hline
\end{tabular}
\end{table*}

\end{document}